\documentclass[twocolumn,superscriptaddress]{revtex4-2} 
\usepackage{graphicx,color}
\usepackage{amsthm}
\usepackage{amsfonts}
\usepackage{algorithmic}
\usepackage{enumerate}
\usepackage{latexsym}
\usepackage{amsmath}
\usepackage{amssymb}
\usepackage{multirow}
\usepackage{array}
\usepackage{diagbox}
\usepackage{subfigure}
\usepackage[colorlinks=true,citecolor=blue,linkcolor=blue]{hyperref}

\usepackage{xcolor}
\usepackage{bm}
\usepackage{braket}

\newcommand{\tck}{\textcolor{black}}

\usepackage{soul}

\def\avg#1{\left\langle#1\right\rangle}

\emergencystretch=\maxdimen
\begin{document}

\title{Charge Stripe Manipulation of Superconducting Pairing Symmetry Transition}
\author{Chao Chen}
\thanks{These authors contributed equally to this work.\\}
\affiliation{Department of Physics, Beijing Normal University, Beijing 100875, China\\}
\affiliation{Beijing Computational Science Research Center, Beijing 100084, China\\}
\author{Peigeng Zhong}
\thanks{These authors contributed equally to this work.\\}
\affiliation{Beijing Computational Science Research Center, Beijing 100084, China\\}
\author{Xuelei Sui}
\affiliation{Beijing Computational Science Research Center, Beijing 100084, China\\}
\author{Runyu Ma}
\affiliation{Department of Physics, Beijing Normal University, Beijing 100875, China\\}
\author{Ying Liang}
\affiliation{Department of Physics, Beijing Normal University, Beijing 100875, China\\}
\affiliation{Key Laboratory of Multiscale Spin Physics, Ministry of Education, Beijing 100875, China\\}
\author{Shijie Hu}
\email{shijiehu@csrc.ac.cn}
\affiliation{Beijing Computational Science Research Center, Beijing 100084, China\\}
\affiliation{Department of Physics, Beijing Normal University, Beijing 100875, China\\}
\author{Tianxing Ma}
\email{txma@bnu.edu.cn}
\affiliation{Department of Physics, Beijing Normal University, Beijing 100875, China\\}
\affiliation{Key Laboratory of Multiscale Spin Physics, Ministry of Education, Beijing 100875, China\\}
\author{Hai-Qing Lin}
\affiliation{Center for Correlated Matter and School of Physics, Zhejiang University, Hangzhou 310058, China\\}
\affiliation{Beijing Computational Science Research Center, Beijing 100084, China\\}
\affiliation{Department of Physics, Beijing Normal University, Beijing 100875, China\\}
\author{Bing Huang}
\email{bing.huang@csrc.ac.cn}
\affiliation{Beijing Computational Science Research Center, Beijing 100084, China\\}
\affiliation{Department of Physics, Beijing Normal University, Beijing 100875, China\\}

\begin{abstract}
Charge stripes have been widely observed in many different types of unconventional superconductors, holding varying periods ($\mathcal{P}$) and intensities. However, a general understanding on the interplay between charge stripes and superconducting properties is still incomplete. Here, using large-scale unbiased numerical simulations on a general inhomogeneous Hubbard model, we discover that the charge-stripe period $\mathcal{P}$, which is variable in different real material systems, could dictate the pairing symmetries --- $d$ wave for $\mathcal{P} \ge 4$, $s$ and $d$ waves for $\mathcal{P} \le 3$. In the latter, tuning hole doping and charge-stripe amplitude can trigger a $d$-$s$ wave transition and magnetic-correlation shift, where the $d$-wave state converts to a pairing-density wave state, competing with the $s$ wave. These interesting phenomena arise from an unusual stripe-induced selection rule of pairing symmetries around on-stripe region and within inter-stripe region, giving rise to a critical point of $\mathcal{P}=3$ for the phase transition. In general, our findings offer new insights into the differences in the superconducting pairing mechanisms across many $\mathcal{P}$-dependent superconducting systems, highlighting the decisive role of charge stripe.
\end{abstract}
\maketitle

{\textbf{Introduction.}} Developing the universal understanding of the intertwisting mechanism between different symmetry-breaking orders is one of the most challenging goals in unconventional superconductors. Initially, the emergence of charge orders in a stripe phase was widely discovered in cuprates,
e.g., La$_2$CuO$_4$\cite{Tranquada1995,Abbamonte2005}, $R$Ba$_2$Cu$_3$O$_6$\cite{science.1223532}, Bi$_2$Sr$_2$CaCu$_2$O$_8$\cite{science.1243479}, and other family materials, sparking significant interest in their origins\cite{Fradkin2012}.
Soon after that, similar charge stripes were later observed in iron-based superconductors, e.g., FeSe\cite{Wang2016}, and Ni-based superconductors, e.g., infinite-layer nickelates\cite{Gu2020, Wang2021, arxiv.2201.12971, arXiv:2201.10038,Ji2023,Cheng2024} and Ruddlesden-Popper-phase nickelates\cite{PhysRevB.61.R854}.
Very recently, the charge stripes were also found in the kagome-lattice superconductors CsV$_3$Sb$_5$\cite{Zheng2022} and CsCr$_3$Sb$_5$\cite{Liu2024}.
Clearly, the widespread existence of charge stripes in variable unconventional superconductors highlights their significant role in relation to superconductivity.
Interestingly, the period $\mathcal{P}$ and intensity $V_0$ of charge stripe, as marked in Fig.~\ref{Fig:phase}(a),
are variable in different materials, which could be tunable by external factors like pressures and defects \cite{Ding2023, arXiv:2306.15086, arXiv:2307.13569, Parzyck2024}, opening potential possibilities to manipulate superconducting pairing symmetry.

\begin{figure}[h]
\includegraphics[scale=0.567, trim = 4 4 5 5, clip]{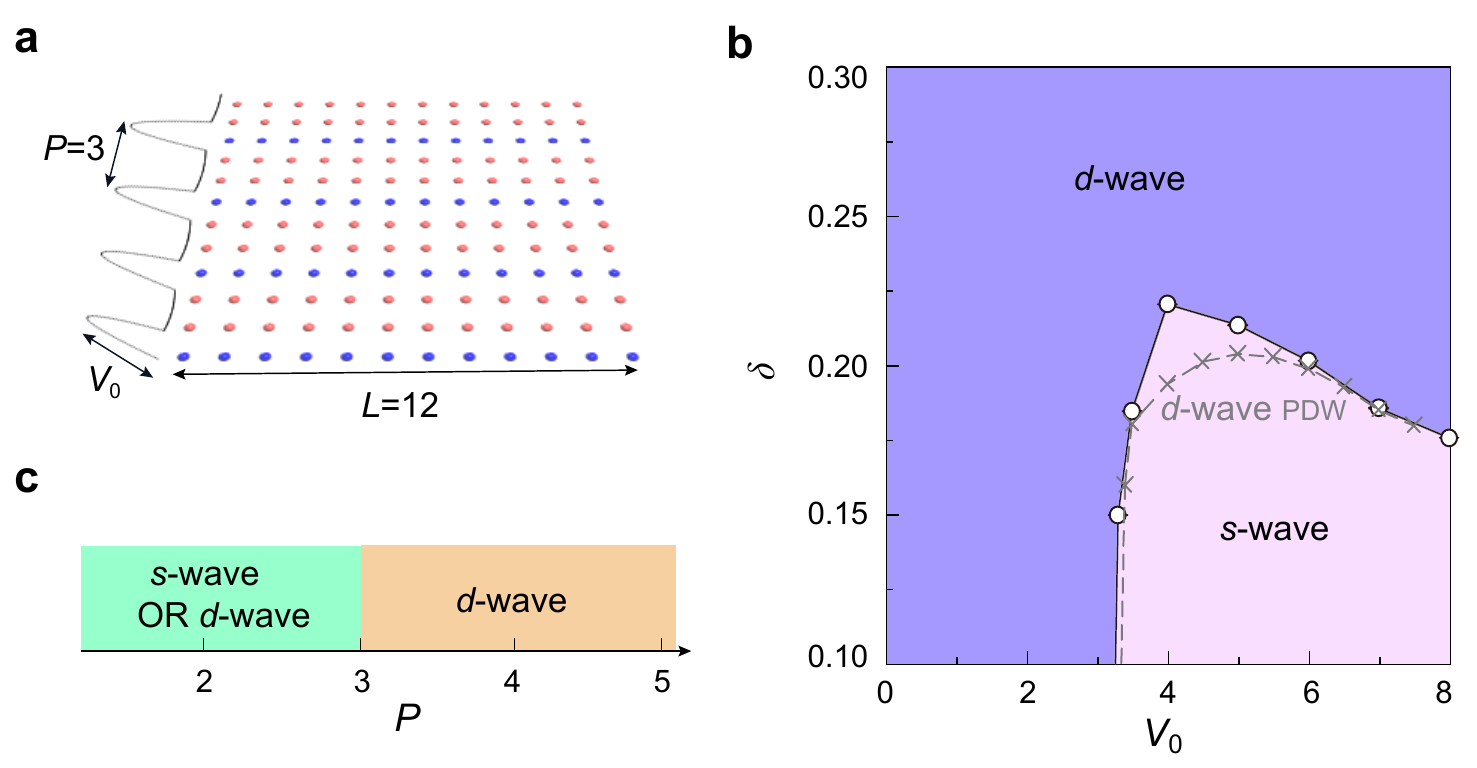}
\caption{(a) Geometry of the square lattice with periodic charge stripes. $\mathcal{P}$ denotes charge-stripe period, $L$ denotes lattice size, and $V_0$ denotes charge-stripe amplitude. Total number of sites is $N=L\times L$, and $L=12$ is fairly large for a square lattice. Blue (red) circles label the site with (without) the inclusion of $V_0$, representing the on-stripe (inter-stripe) region. (b) DQMC-calculated phase diagram of the inhomogeneous Hubbard model with $\mathcal{P}=3$, on-site repulsion strength $U/t=4$, and temperature $T=t/5$. $\delta$ represents hole-doping concentration. Note that the $d$-$s$ wave transition is observed even at zero temperature based on DMRG simulations. Phase boundary of solid-line is determined by effective pairing strength $\bar{P}_{\alpha}$ at each ($V_0$, $\delta$). Dashed-line denotes the region where $d$-wave state is transformed into PDW state, competing with $s$-wave state. Note that $s$-wave state is always more stable than PDW state. (c) Dominant pairing symmetry depends on $\mathcal{P}$, where $\mathcal{P}=3$ is a critical point.
}
\label{Fig:phase}
\end{figure}

Since its inception, the Hubbard model has served as an archetypal model for elucidating strongly correlated phenomena\cite{Keimer2015, annurev102024}.
Despite its simplicity, it can uncover rich quantum phases in condensed matter physics~\cite{RevModPhys.66.763, RevModPhys.84.1383, RevModPhys.87.457, science.aam7127, science.aak9546, PhysRevB.35.3359, PhysRevB.62.R9283}. 
For example, the Hubbard model under different conditions can effectively capture $d$-wave pairing symmetry~\cite{RevModPhys.66.763,RevModPhys.84.1383},
stripe order~\cite{RevModPhys.87.457,science.aam7127,science.aak9546,npjQM.3.22}, and antiferromagnetic (AFM) order~\cite{PhysRevB.62.R9283,PhysRevLett.94.156404} in cuprate-like square lattices, and it can also demonstrate the interplay between these symmetry-breaking orders~\cite{PhysRevLett.104.247001}.
Interestingly, previous studies have suggested that the spontaneous formation of charge stripe in a square lattice could be sensitive to variations in model parameters and lattice boundary conditions~\cite{science.adh7691}.
Alternatively, the charge stripes can be artificially induced as external fields to explore its relationship with superconductivity. For example, in square-lattice models with $\mathcal{P}=4$ for simulating cuprates, an enhancement of $d$-wave pairing symmetry is observed~\cite{PhysRevB.86.184506,pnas.2109406119,PhysRevB.72.060502} over a broad range of $V_0$, which can be attributed to the intensified AFM correlations between the stripes, accompanied by a $\pi$-phase shift in the system~\cite{PhysRevB.86.184506}.
Until now, however, a comprehensive understanding of the interplay between charge stripe, varying $\mathcal{P}$ and $V_0$ values, and superconducting pairing symmetry remains lacking, which may prevent a deeper insight into the distinct paring symmetries observed across different systems.

The unbiased determinant quantum Monte Carlo (DQMC) and density-matrix renormalization group (DMRG) methods are widely recognized as two highly accurate and complementary approaches to solve the Hubbard model~\cite{RevModPhys.66.763, RevModPhys.84.1383,science.aam7127}.
While DQMC can effectively capture the trend of physical quantities at finite temperatures, DMRG is powerful in determining them in the ground state.
Here, by combining unbiased DQMC and DMRG simulations on an inhomogeneous square lattice, we discover that the existence of charge stripes with different periods $\mathcal{P}$ [defined in Fig.~\ref{Fig:phase}(a)] plays a very unexpected role in determining the pairing-symmetry transition.
While the $d$-wave is always dominant for $\mathcal{P} \geq 4$, both $s$ (note that this is an extended $s$-wave state afterwards) and $d$-waves can appear when $\mathcal{P} \leq 3$. Taking $\mathcal{P} = 3$ as an example, we discover that the interplay between the hole-doping concentration $\delta$ and charge-stripe amplitude $V_0$ can realize a remarkable $d$-$s$ wave transition in a large region of the phase diagram, in which the critical $V_0$ ($V_0,c$) for the phase transition exhibits a nearly linear dependence of the on-site electron-electron repulsion strength $U$.
The DMRG simulations further reveal that the charge-stripe-induced domain wall can generate an interesting selection rule to produce $s$ and $d$-waves around the on-stripe region and inside the inter-stripe region, respectively.
Therefore, the smaller the $\mathcal{P}$, the stronger the $s$-wave in the system. Accompanying the $d$-$s$ wave transition, there is a novel magnetic-correlation transition, weakening the AFM correlation. These results strongly indicate an inherent interplay between charge stripes, superconducting pairing, and magnetic correlation in the $\mathcal{P}$-dependent systems, in which charge stripes play a vital role in forming the $d$-$s$ wave transition.

{\textbf{$\mathcal{P}$-dependent $d$-$s$ wave transition.}} In the following, we will mainly discuss the model system with charge stripes at $\mathcal{P}=3$ in a minimal single-$d_{{x^2}-{y^2}}$-band Hubbard model, because this simplified model could capture the most intrinsic feature between $\mathcal{P}$ and pairing symmetry and also because a similar dominated role of single-$d_{{x^2}-{y^2}}$-band was observed in cuprates and nickelates~\cite{10.1093nwae194,arxiv.2403.07344}. As shown in Fig.~\ref{Fig:phase}(b), we have systematically calculated the pairing-symmetry diagram as a function of $\delta$ and $V_0$. Here, $\delta$ is set to the range of $0.1\sim0.3$~\cite{PhysRevLett.125.147003, PhysRevLett.125.027001}, and $V_0$ is set to the range of $0\sim8$ based on the realistic situations. For example, the $V_0$ induced by variable valence Ni charge-state in the stripe of infinite-layer nickelates is estimated to be $\sim$6, which is further tunable under the external conditions~\cite{arXiv:2306.15086,arXiv:2307.13569,Parzyck2024}. When $V_0$ is larger than a critical value of $V_{0,c}\sim$3.25, there is a clear pairing-symmetry transition from $d$ to $s$ waves in a large $\delta$ range of $0.1\sim0.23$. As will be shown later, this $d$-$s$ wave transition is robust against different $U/t$ and $T/t$ values. For comparison, we have also calculated the cases of $\mathcal{P}=2$ and $\mathcal{P}=4$. Interestingly, when  $\mathcal{P}=2$, a similar $d$-$s$ wave transition can be observed at an even smaller $V_{0,c}$ with a much sharper transition slope (Fig.~S1). On the other hand, when $\mathcal{P}=4$, only $d$-wave is observed and $d$-$s$ wave transition cannot exist in the same $\delta$ range (Fig.~S2). As summarized in Fig.~\ref{Fig:phase}(c), these calculations lead us to an interesting conclusion that $\mathcal{P}=3$ is a critical point for the pairing-symmetry transition, that is, the $\delta$/$V_0$-dependent $d$-$s$ wave transition can only exist when $\mathcal{P} \leq 3$. Importantly, this finding is regardless of whether it is a single-band or multi-band model (Fig.~S3),  being a general feature in $\mathcal{P}$-dependent superconducting systems.

\begin{figure*}[tph]
\includegraphics[scale=0.4,trim = 5 43 5 43, clip]{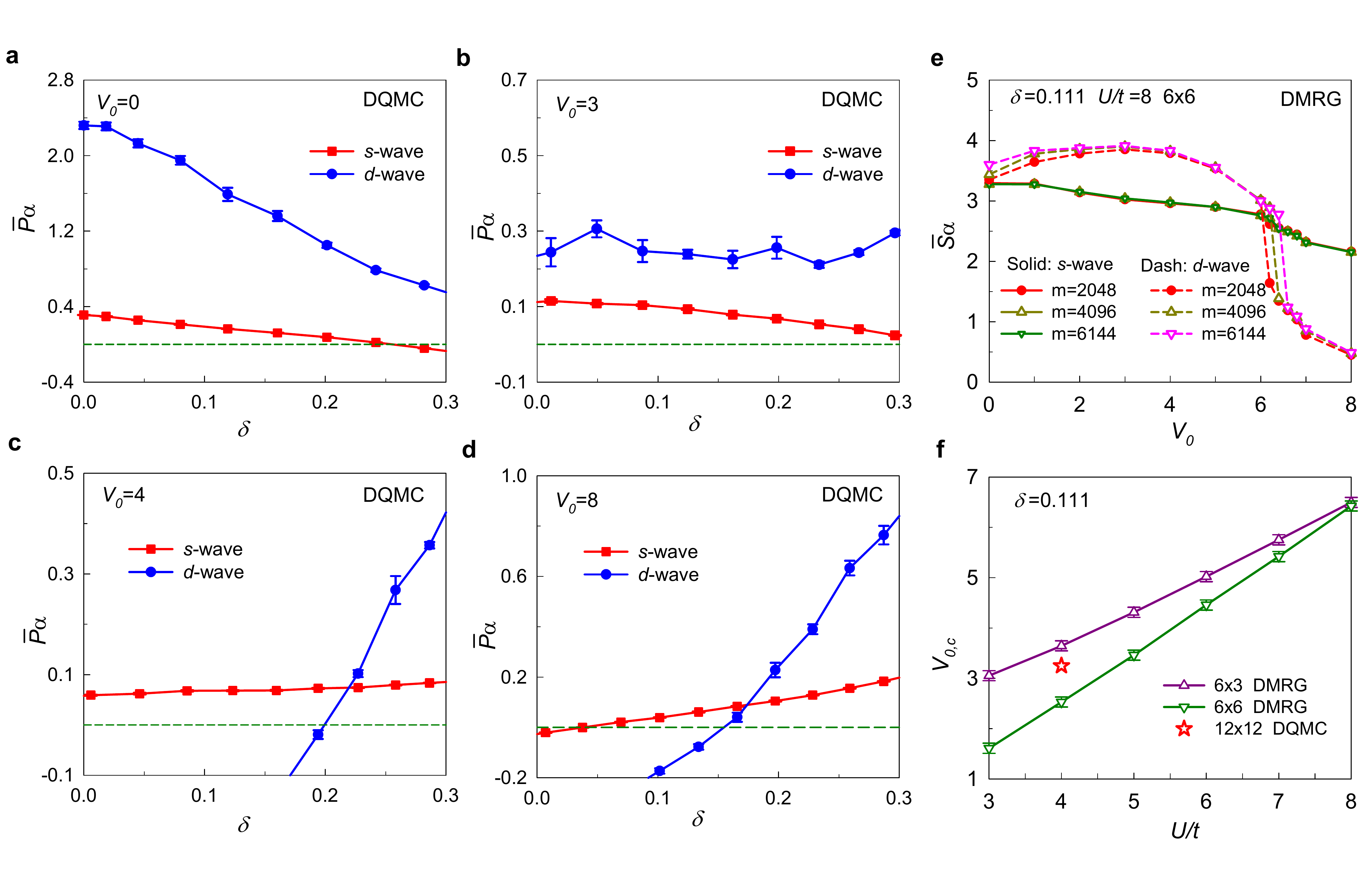}
\caption{DQMC-calculated $\bar{P}_{\alpha}$ as a function of $\delta$ at $T=t/5$ and $U/t=4$ with $\mathcal{P}=3$ on a $L=12$ lattice for (a) $V_0$=0, (b) $V_0$=3, (c) $V_0$=4, and (d) $V_0$=8. (e) DMRG-calculated effective zero-momentum pair-pair structure factor $\bar{S}_{\alpha}$ as a function of $V_0$ at $\delta$=0.111 and $U/t=8$ on the $6\times6$ cylinder. While a small discrepancy is observed, it will not affect our conclusion within small error bars. (f) DMRG-calculated $V_{0,c}$ for $d$-$s$ wave transition with different $U$ on both $6\times3$ and $6\times6$ cylinders, exhibiting a nearly linear function of $U$. For comparison, DQMC result on a $12\times12$ lattice at $U/t=4$ and $T=t/5$ is also marked here.
}
\label{Fig:spin}
\end{figure*}

To clearly understand the role of $V_0$ in the $d$-$s$ wave transition, we have plotted the $\delta$-dependence of effective pairing interaction $\bar{P}_{\alpha}$ with the typical parameters of $T=t/5$ and $U/t=4$ under different $V_0$. As shown in Fig.~\ref{Fig:spin}(a), without charge stripes ($V_0=0$), $\bar{P}_{d}$, which is strongest at $\delta=0$, is robust and more stable than that of $\bar{P}_s$ at different $\delta$. Meanwhile, the $s$-wave pairing is suppressed ($\bar{P}_{s}<0$) at large $\delta$. As shown in Fig.~\ref{Fig:spin}(b), when $V_0$=3, $\bar{P}_{d}$ is rapidly decreased in a much faster way than that of $\bar{P}_{s}$. This indicates that $s$-wave pairing is more robust against the charge-stripe potential compared to $d$-wave pairing. Importantly, as shown in Fig.~\ref{Fig:spin}(c), when $V_0$=4, $\bar{P}_{s}$ eventually becomes more stable than $\bar{P}_{d}$ over an extensive $\delta$ range ($0<\delta \leq0.22$), leading to a remarkable $d$-$s$ wave transition. In particular, $d$-wave pairing is fully suppressed at $0<\delta \leq0.2$ under $V_0=4$, eventually {transformed into} a $d$-wave PDW state to compete with $s$-wave state, as discussed later. As shown in Fig.~\ref{Fig:spin}(d), when $V_0=8$, $\bar{P}_s$ maintains more stable than $\bar{P}_d$ in the moderate $\delta$ ($0.05< \delta \leq0.15$). However, for sufficiently large $\delta$, $\bar{P}_{d}$ is always more stable than $\bar{P}_{s}$, regardless of the $V_0$, as also shown in the phase diagram of Fig.~\ref{Fig:phase}(b).

The above finite-temperature DQMC conclusion holds at a much lower temperature of $T$=$t$/12 (Fig.~S4). To further confirm the ground-state properties at zero temperature, we have systematically calculated effective zero-momentum pair-pair structure factor $\bar{S}_{\alpha}$ using DMRG method with different cylinders and $U$ (Fig.~S5). For example, Fig.~\ref{Fig:spin}(e) shows $\bar{S}_{\alpha}$ as a function of $V_0$ at $\delta$=0.111 and $U/t=8$ on a 6$\times$6 cylinder. $\bar{S}_{d}$ is dominant when $V_0$ is smaller than $\sim6.2$. Interestingly, when $V_0$ is bigger than $\sim6.2$, $\bar{S}_{s}$ becomes more robust.
Therefore, at ground state, charge inhomogeneity can also support a remarkable $d$-$s$ wave transition, demonstrating that the finite-temperature trend obtained from DQMC simulations is reliable at zero temperature. In Fig.~\ref{Fig:spin}(f), we have plotted $V_{0,c}$ as a function of  $U$ for the observed pairing-symmetry transition with two cylinders. Remarkably, $V_{0,c}$ displays a nearly linear relationship with $U$ for both $6\times3$ and $6\times6$ cylinders. As $U$ increases, so does $V_{0,c}$, providing a guideline for understanding or manipulating the pairing-symmetry transition. For a typical $U/t$=4, the DMRG-calculated $V_{0,c}$ in $6\times3$ and $6\times6$ cylinders exhibit slightly different values indicating the lattice-size dependence. However, these values are overall consistent with DQMC results. Therefore, our results undisputedly demonstrate that this $d$-$s$ wave transition exists in a $\mathcal{P}=3$ system and that the $V_{0,c}$ depends on $U$.

We have further investigated the critical role of different parameters on $\bar{P}_{\alpha}$. Here, we choose the cases of $\delta=0.3$ ($d$-wave-dominated region) and $\delta=0.18$ ($s$-wave-dominated region). Figures~\ref{Fig:FigP}(a)-(b) show the case of $d$-wave pairing at $\delta=0.3$. In Fig.~\ref{Fig:FigP}(a), we calculate the temperature-dependent $\bar{P}_{d}$ for different $V_0$. As temperature is lowered, $\bar{P}_{d}$ increases rapidly.  Importantly, it is observed that $d$-wave pairing is enhanced with the increase of $V_0$, indicating the important role of charge fluctuation \cite{PhysRevB.86.184506, PhysRevLett.104.247001}. This enhancement may be caused by the appearance of more nearly half-filled inter-stripe regions for larger $V_0$ at $\delta=0.3$ (Fig.~S6). On the other hand, Fig.~\ref{Fig:FigP}(b) shows that the $\bar{P}_{d}$ is enhanced by larger $U$, suggesting the importance of electron-electron correlation. Importantly, the lattice size effect of $\bar{P}_{d}$ is weak, i.e., $L=9$, $12$, and $15$ exhibit almost identical results.

\begin{figure}[tbp]
\includegraphics[scale=0.283,trim = 170 92 20 0, clip]{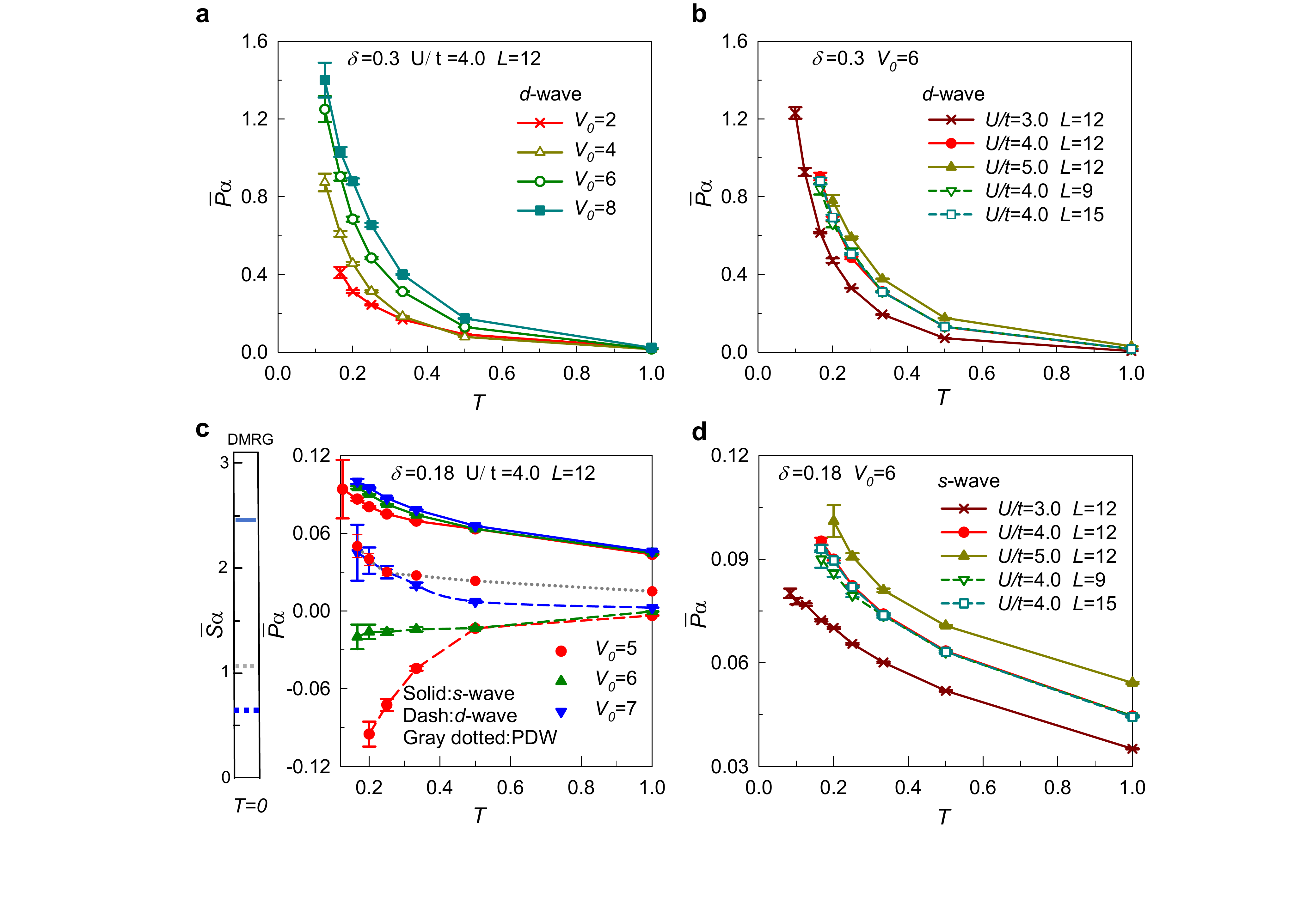}
\caption{DQMC-calculated $\bar{P}_{\alpha}$ as a function of temperature (a) for
different $V_0$ at $\delta=0.3$, $U/t=4$, and $L=12$,
(b) for different $U/t$ or $L$ at $\delta=0.3$ and $V_0=6$. (c)-(d) are similar to (a)-(b) but for the cases of $\delta=0.18$. Gray-dotted line in (c) represents the $d$-wave PDW at $V_0=5$. Left of (c): DMRG-calculated $\bar{S}_{\alpha}$ of $s$-wave (blue-solid) and $d$-wave (blue-dash), and the peak value of $d$-wave PDW $\bar{S}_d(\mathbf{q})$ (gray-dash) at $\delta=0.111$, $U/t=8$, $V_0=7$ on a 6$\times$6 cylinder.
}
\label{Fig:FigP}
\end{figure}

Figures~\ref{Fig:FigP}(c)-(d) show the case of $s$-wave pairing at $\delta=0.18$. In Fig.~\ref{Fig:FigP}(c), we present the temperature dependence of $\bar{P}_{s}$, in which $\bar{P}_{d}$ is also plotted here for comparison. For $V_0=5\sim7$, $\bar{P}_{s}$ is positive and increases slowly with decreasing temperature. The larger $V_0$, the stronger $\bar{P}_{s}$. However, $\bar{P}_{d}$ is negative at $V_0=5\sim6$ and becomes positive at $V_0=7$. So, below $V_0=7$, $\bar{P}_{d}$ is less stable than $\bar{P}_{s}$ at all the considered temperature ranges. It is curious to understand the origin of the suppression of $d$-wave state, which suggests that there may be an unusual phase transition. To confirm our speculation, we have systematically calculated the possible PDW state in $\mathcal{P}=3$ system. Taking $V_0=5$ as an example [Fig.~\ref{Fig:FigP}(c)], interestingly, the peak of $P_{d}^\text{PDW}(\mathbf{q})$ moves away from zero momentum and the system shows a tendency to form a PDW state.  Although $P_{d}^\text{PDW}(\mathbf{q})$ is positive, it is still less stable than $\bar{P}_{s}$. In addition, we further calculate the competition between $P_{d}^\text{PDW}(\mathbf{q})$ and $\bar{P}_{s}$ under different $V_0$ and $\delta$ (Fig.~S7), and find that $s$-wave state is always more stable than PDW state. This may account for the challenge to observe PDW in nickelates, whch is hidden behind the $s$-wave. In the phase diagram of Fig.~\ref{Fig:phase}(b), we have also plotted the boundary where PDW states emerge, which is close to the boundary of $s$-wave state. To further confirm our DQMC conclusion, we have plotted DMRG-calculated $\bar{S}_\alpha$ of the $s$- and $d$-waves, and the peak value of PDW $\bar{S}_d(\mathbf{q})$ at $\delta=0.111$, $U/t=8$, $V_0=7$ on a $6\times6$ cylinder, supporting the dominance of $s$-wave at zero temperature (see more cases in Fig.~S8).
Fig.~\ref{Fig:FigP}(d) shows $\bar{P}_{s}$ as a function of temperature at different $U$ and $L$. Similar to that in Fig.~\ref{Fig:FigP}(b), it is obvious that $\bar{P}_{s}$ is also enhanced with increasing $U$ and shows a very weak lattice-size effect. 
Furthermore, our constrained path quantum Monte Carlo (CPQMC) and DMRG simulations also suggest the possible emergence of long-range $s$-wave superconducting order within the investigated parameter region (Fig.~S9 and ~S10).

\begin{figure}[tbp]
\includegraphics[scale=0.3,trim = 80 64 100 51, clip]{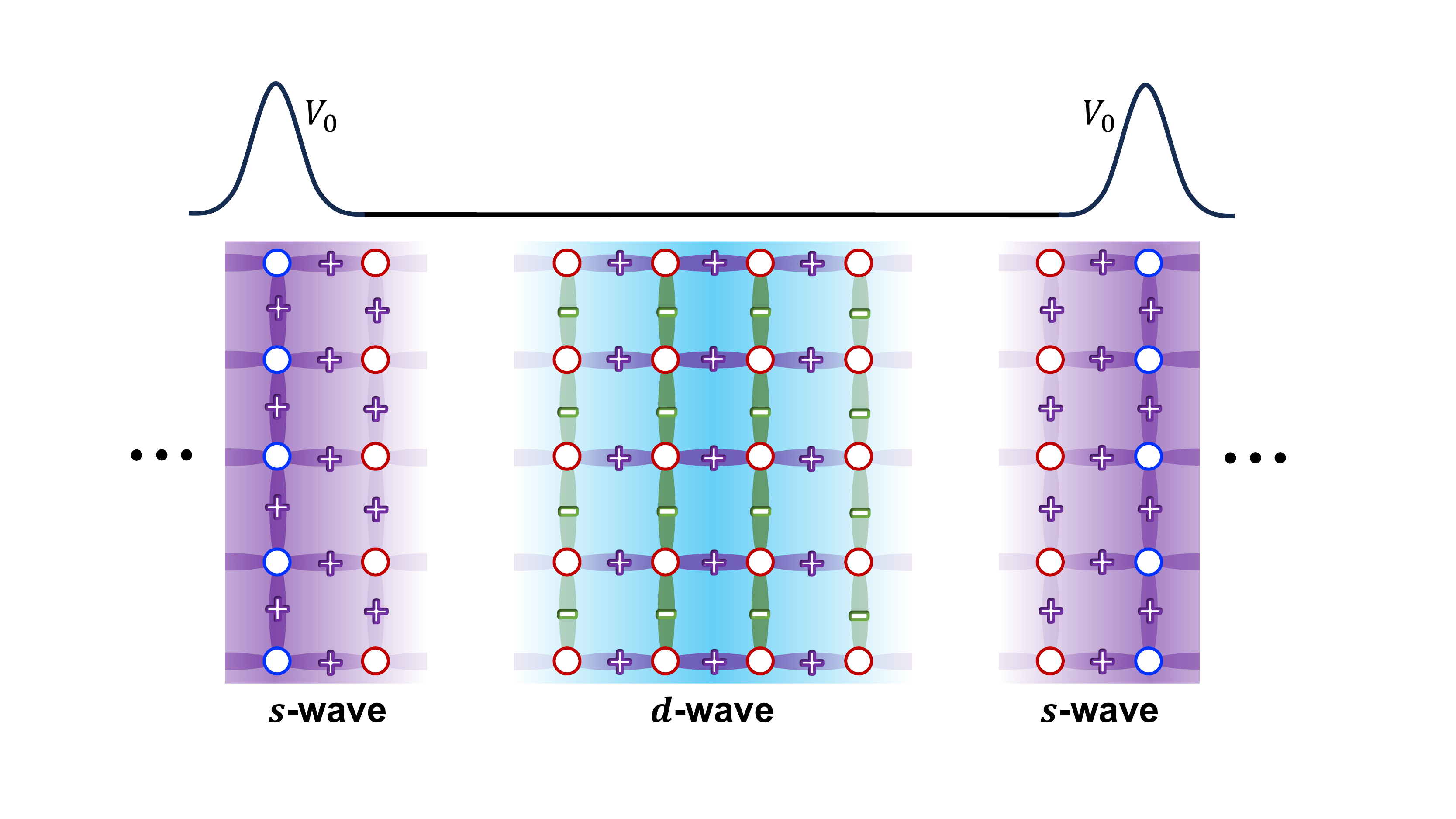}
\caption{Sketch depicts the $d$-$s$ wave transition by analyzing the condensate wave function of the dominant Cooper pair mode $\zeta_0 (\mathbf{i}\bm{\delta}_l)$ based on DMRG simulations. Blue (red) circles label the sites in the on-stripe (inter-stripe) region with (without) $V_0$. Purple (green) bonds indicate positive (negative) values of $\zeta_0 (\mathbf{i}\bm{\delta}_l)$. Symmetric $s$-wave patterns only occur near on-stripe regions (domain-walls) at moderate $V_0$ and $\delta$, i.e., horizontal and vertical bonds have the same signs. On the contrary, inter-stripe region always benefits asymmetric $d$-wave, i.e., horizontal and vertical bonds have opposite signs. Due to the competition between the $d$- and $s$-wave pairing symmetries, global $s$-wave pattern can be stabilized with $\mathcal{P}\le 3$.
}
\label{Fig:reason}
\end{figure}

{\textbf{Origin of $d$-$s$ wave transition.}} It is interesting to understand the physics insight behind this $d$-$s$ pairing-symmetry transition. In Fig.~\ref{Fig:reason}, based on the ground-state DMRG analysis on the condensate wave function, we realize that this phase transition is strongly related to charge-stripe-induced potential fluctuation, where the domain-walls can form around the on-stripe region (blue-circle in Fig.~\ref{Fig:reason}). Specifically, the DMRG-calculated dominant Cooper pair mode $\zeta_0 (\mathbf{i} \bm{\delta}_l)$ supports that a clear local pattern of $s$-wave pairing can emerge around on-stripe regions at moderate $V_0$ and $\delta$, regardless of  $\mathcal{P}$, where horizontal and vertical bonds have the same signs (Fig.~S11). On the contrary, inter-stripe region (red-circle in Fig.~\ref{Fig:reason}) is always beneficial to asymmetric $d$-wave, as long as $\mathcal{P}$ is sufficiently large, where horizontal and vertical bonds exhibit opposite signs (Fig.~S11). In brief, without the domain-wall, the system favors asymmetric $d$-wave patterns. In the presence of domain-walls, the influence of domain-walls on pairing symmetry is local, and $s$-wave patterns can only be prominent near on-stripe region at moderate $V_0$ and $\delta$. The smaller $\mathcal{P}$, the more the $s$-wave components can be generated in the system. When $\mathcal{P} \ge 4$, the inter-stripe $d$-wave region plays a dominant role in forming global $d$-wave pairing in the system. However, when $\mathcal{P}\le 3$, the intensity of $s$-wave pattern near on-stripe region is sufficiently strong to convert the global pairing symmetry from $d$ to $s$. This understanding not only can explain why the $d$-$s$ wave transition is more accessible in a smaller $\mathcal{P}$ system [Fig.~S1 and Fig.~\ref{Fig:phase}(c)], but also suggests that the local $s$-wave pairing may also exist in $\mathcal{P} \ge 4$ ($d$-wave-dominant) systems, as long as the $V_0$ and $\delta$ are in a suitable region.

\begin{figure}[tbp]
\includegraphics[scale=0.527,trim = 215 32 130 20, clip]{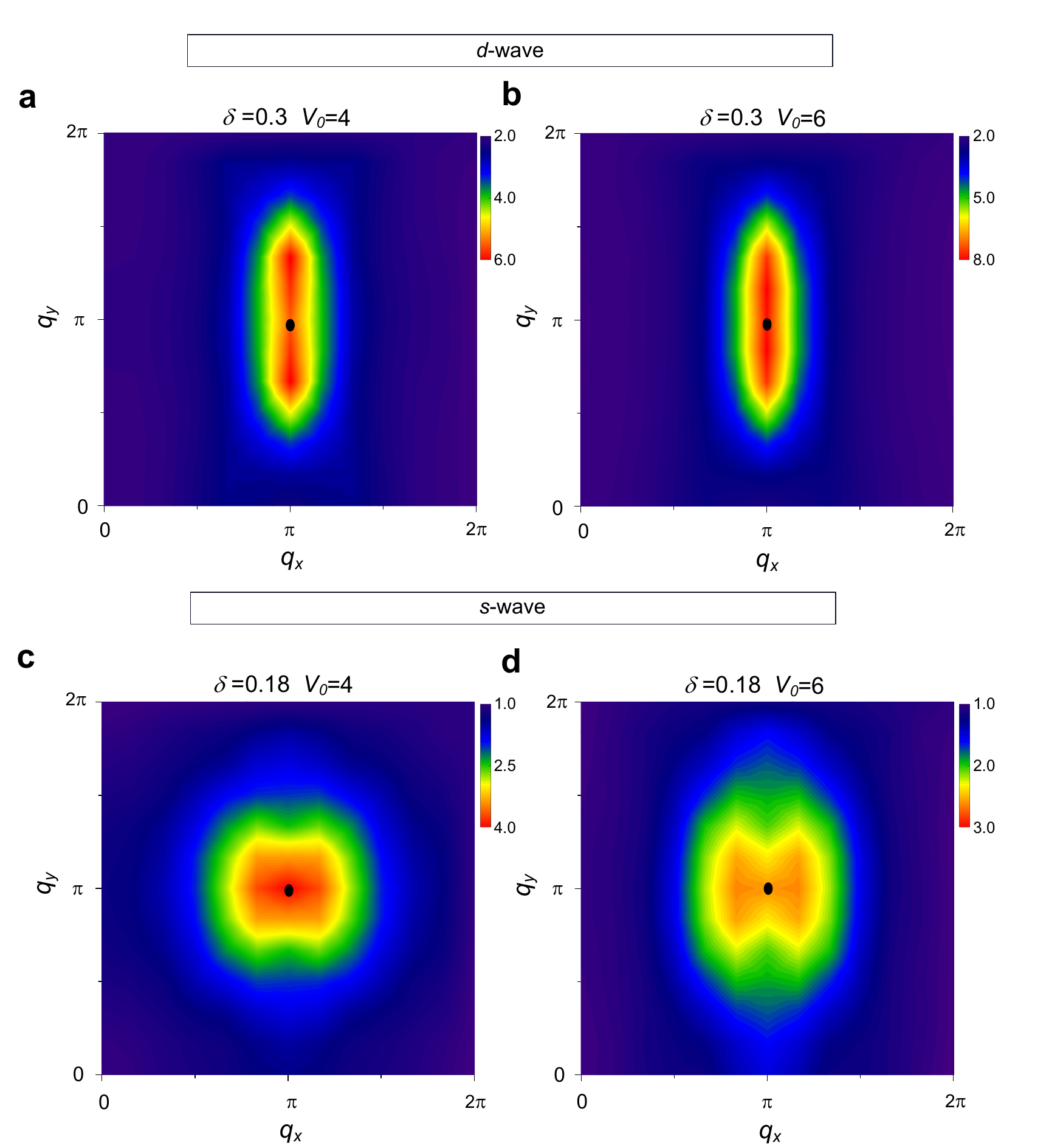}
\caption{DQMC-calculated spin susceptibility $\chi_{s}(\mathbf{q})$ in the first Brillouin zone at $T=t/5$, $U/t=4$ with (a) $\delta=0.3$ and $V_0$=4, (b) $\delta=0.3$ and $V_0$=6, (c) $\delta=0.18$ and $V_0$=4, and (d) $\delta=0.18$ and $V_0$=6. Here, (a)-(b) are for the $d$-wave cases, and (c)-(d) are for the $s$-wave cases.
}
\label{Fig:sus0.7}
\end{figure}

Besides the pairing-symmetry transition, it is also curious to understand the role of charge stripe on the modulation of spin susceptibility [$\chi_{s}(\mathbf{q})$].
In Figs.~\ref{Fig:sus0.7}(a) and (b), we have calculated the $\chi_{s}(\mathbf{q})$ for two different $V_0$ at $\delta=0.3$ in the $\mathbf{q}$-space (see more $V_0$ cases in Fig.~S12). In the $d$-wave region, the system behaves as the AFM correlation. One can see that the $(\pi,\pi)$ magnetic correlation is enhanced as the $V_0$ increases, i.e., the system exhibits a stronger AFM fluctuation along the direction of stripes ($x$ direction) with larger $V_0$. This AFM-correlation enhancement is possibly caused by more nearly half-filled inter-stripe regions (Fig.~S6), similar to the behavior of enhanced $d$ pairing symmetry at $\delta=0.3$.
Moreover, sub-peaks emerge at $q_y = \pi \pm \pi / \mathcal{P} = 2\pi/3$ and $4\pi/3$, reflecting the incommensurate spin correlations observed in the $d$-wave superconductor~\cite{npjQM.3.22}, and are gradually supressed as increasing $V_0$ from $4$ to $6$.

The case is dramatically changed in the $s$-wave region. In Figs.~\ref{Fig:sus0.7}(c) and (d), we have calculated $\chi_{s}(\mathbf{q})$ at $\delta=0.18$ with two different $V_0$, in which the $s$-wave pairing symmetry is dominated. Surprisingly, along with the $d$-$s$ wave transition, the AFM correlation at $(\pi,\pi)$ is weakened. In detail, $\chi_{s}(\mathbf{q})$ shows a dumbbell shape, different from the rod shape for the $d$-wave at $\delta=0.3$. The dumbbell distribution becomes more obvious as $V_0$ increases (see more $V_0$ cases in Fig.~S13).
Besides, the $(\pi,\pi)$ magnetic correlation and dominant pairing correlation exhibit a very similar temperature dependence (Fig.~S14). Given that magnetism and superconductivity simultaneously exhibit dramatical differences in these two doping cases of $\delta=0.3$ and $0.18$, it indicates that the pairing-symmetry transition and magnetic-correlation transition are strongly interwoven.

{\textbf{Outlook.}}
Both charge and spin stripes are widely
observed in many superconductors.  Although
the spin stripe itself is interesting in a model study, it is
beyond the focus for our current study. Meanwhile, although the major conclusion is described by a minimal Hubbard model, it is robust against the multi-band model (Fig.~S3) or different stripe styles (Fig.~S15). Since the charge stripe in a real material system might be tunable under some external conditions, combined with the linear relationship between $V_{0,c}$ and $U$, our study provides a novel idea of charge-stripe engineering of pairing symmetry. During the $d$-$s$ wave phase transition, the competition between PDW and $s$-wave provides an important opportunity to explore the exotic intertwining phenomenon between PDW, $d$-wave, and $s$-wave.

{\textbf{Model and Method}}
The two-dimensional Hubbard Hamiltonian on a square lattice with nearest-neighbor hopping $t$ and Coulomb repulsion $U$ is written as
\begin{eqnarray}
\label{Hamiltonian}
\hat H=&-&t\sum_{{\langle\bf i}, {\bf j} \rangle \sigma}  (c_{{\bf i}\sigma}^\dagger c_{{\bf j} \sigma}^{\phantom{\dagger}}+c_{{\bf j}\sigma}^\dagger c_{{\bf i} \sigma}^{\phantom{\dagger}})
+U \sum_{{\bf i}}   n_{{\bf i}\uparrow}  n_{{\bf i}\downarrow} \nonumber \\
&-& \mu \sum_{\bf i}  (n_{{\bf i}\uparrow}+n_{{\bf i}\downarrow}) +V_0\sum_{mod(i_y,\mathcal{P})=0} (n_{{\bf i}\uparrow}+n_{{\bf i}\downarrow}) .
\end{eqnarray}
Here, $c_{\mathbf{i}\sigma}$ ($c_{\mathbf{i}\sigma}^{\dag}$) annihilates (creates) electrons at site $\mathbf{i}$ with spin $\sigma$ ($\sigma=\uparrow$, $\downarrow$), and $n_{\mathbf{i}\sigma}=c_{\mathbf{i}\sigma}^{\dag}c_{\mathbf{i}\sigma}^{\phantom{\dag}}$ is the particle number operator for the electron.
We set the nearest-neighbor hopping $t=1$ as the energy unit. $\mu$ is a global chemical potential for all sites, and $V_0$ is an additional potential exerted on a set of on-stripe rows $\mathbf{i}=(i_x,\ i_y)$ where $i_y=0$ modulo $\mathcal{P}$, that is, $\text{mod}(i_y, \mathcal{P})=0$. The larger $V_0$, the stronger the charge fluctuation. Accordingly, as shown in Fig.~\ref{Fig:phase}(a), the charge stripe with tunable oscillation strength can be introduced externally via a raised energy $V_0$. To further confirm our results, we have also selected a cosine-like varying modulation (Fig.~S15). Interestingly, $V_{0,c}$ for the $d$-$s$ wave transition becomes even smaller when we choose the cosine-like varying charge modulation.

We note that the purpose of this model is not to address the origin of the stripe formation, as this is still an open question. Instead, it allows us to estimate the characteristics of spin and pairing correlations in the presence of pre-existing charge orders. This is an appropriate approximate model when the energy scale of the stripe formation is greater than that of superconductivity~\cite{PhysRevB.72.060502, PhysRevLett.104.247001, PhysRevB.86.184506, Wu2011, PhysRevX.6.021004, PhysRevLett.125.147003, PhysRevLett.125.027001}.

{\it{DQMC method:}} Our calculations are mainly performed on the lattice shown in Fig.~\ref{Fig:phase}(a) using the DQMC method with periodic boundary conditions. This unbiased numerical method is powerful and reliable to investigate strongly-correlated electrons~\cite{PhysRevD.24.2278, PhysRevB.40.506, PhysRevLett.120.116601, PhysRevB.99.195147, science.abg9299}. The basic strategy of the finite-temperature DQMC method is to express the partition function $Z$=$\mathrm{Tr}\exp(-\beta H)$ as a high-dimensional integral over a set of random auxiliary fields. The integration is then accomplished by Monte Carlo sampling.
In our DQMC simulations, 8,000 warm-up sweeps are conducted to equilibrate the system, and an additional $10,000\sim1,200,000$ sweeps are performed for measurements, which are divided into $10\sim20$ bins. Besides, two local updates are performed between measurements.
In the process of eliminating the on-site interaction, the inverse temperature $\beta=1/T$ is discretized. And the discretization mesh $\Delta\tau=0.1$ of $\beta$ is chosen small enough so that the resulting Trotter errors are typically smaller than those associated with the statistical sampling.

We have performed a systematical analysis of the infamous sign problem~\cite{science.abg9299} in our DQMC simulations.
The average sign decreases quickly as the inverse temperature exceeds 3, and the sign problem gets worse for higher $U$ and larger $L$. In most of our calculations, the average sign keeps as $>$0.55 (see Fig.~S16). In order to explore the lower temperature behavior of $\bar{P}_{\alpha}$, the average sign keeps as $>$0.4 (Fig.~S17). In short, the conclusions obtained from our DQMC calculations are reliable.

To explore the effects of the charge-density modulation on superconductivity, we define the pairing interaction as
\begin{equation}
P_\alpha = \frac{1}{N_s} \sum_{\mathbf{i},\mathbf{j}} \mathcal{D}_\alpha (\mathbf{i}, \mathbf{j})\ ,
\label{shi29}
\end{equation}
where
\begin{equation}
\mathcal{D}_\alpha (\mathbf{i}, \mathbf{j}) = \int_0^\beta d \tau \left\langle \Delta
_\alpha^\dag (\mathbf{i},\tau) \Delta_\alpha^{\phantom{\dag}} (\mathbf{j},0) \right\rangle_T
\end{equation}
gives the zero-frequency pair-pair correlation function between sites $\mathbf{i}$ and $\mathbf{j}$, $\alpha$ represents the pairing symmetry,
the corresponding order parameter $\Delta_\alpha (\mathbf{i},\tau) = e^{H\tau} \Delta_\alpha (\mathbf{i},0) e^{-H\tau}$ and $\Delta_\alpha^\dag (\mathbf{i},0)$ is written as
\begin{equation}
\Delta_\alpha^\dag (\mathbf{i},0)\ = \sum_l f_\alpha ^{*} (\bm{\delta}_l) \mathcal{C}_{\mathbf{i} \bm{\delta}_l}^\dag
\label{shi22}
\end{equation}
with $\mathcal{C}_{\mathbf{i} \bm{\delta}_l} = c_{\mathbf{i} \uparrow} c_{\mathbf{i}+\bm{\delta}_l \downarrow} - c_{\mathbf{i} \downarrow} c_{\mathbf{i} + \bm{\delta}_l \uparrow}$ denoting the operator for the Cooper pair on the sites $\mathbf{i}$ and $\mathbf{i} + \bm{\delta}_l$, and $f_\alpha (\bm{\delta}_l)$ stands for the form factor of pairing function. The vectors $\bm{\delta}_l$ ($l=1$, $2$, $3$, $4$) denote the nearest-neighbor connections, and $\bm{\delta}_l$ is $\pm \hat{x}$ and $\pm \hat{y}$. Considering the structure of the square lattice, the possible singlet pairing forms are given by either the extended $s$-wave or the $d$-wave, which have the following form factor~\cite{PhysRevB.39.839, PhysRevB.86.184506},
\begin{eqnarray}
\begin{split}
&\text{$s$-wave} : f_s (\bm{\delta}_l)=+1\ ,\\
&\text{$d$-wave} : f_d (\bm{\delta}_l)=\left\{
\begin{array}{c}
+1\ \text{for}\ {\bm{\delta}_l}=\pm \hat{x}\\
-1\ \text{for}\ {\bm{\delta}_l}=\pm \hat{y}
\end{array}\ ,
\right.
\end{split}
\label{shi33}
\end{eqnarray}
In practice, the effective pairing interaction $\bar{P}_{\alpha}$ is a more direct probe to identify the dominant superconducting pairing form \cite{PhysRevLett.110.107002, HUANG2019310}. In order to obtain $\bar{P}_\alpha$, the uncorrelated single-particle contribution $\widetilde{\mathcal{D}}_\alpha (\mathbf{i},\mathbf{j})$ is also calculated, which is reached by replacing $\langle c_{\mathbf{i} \downarrow}^\dag c_{\mathbf{j}\downarrow}^{\phantom{\dag}} c_{\mathbf{i}+\bm{\delta}_l \uparrow}^\dag c_{\mathbf{j} + \bm{\delta}_{l'} \uparrow}^{\phantom{\dag}} \rangle$ in Eq.~\eqref{shi29} with $\braket{c_{\mathbf{i}\downarrow }^\dag c_{\mathbf{j}\downarrow}^{\phantom{\dag}}} \braket{c_{\mathbf{i}+\bm{\delta}_l \uparrow}^\dag
c_{\mathbf{j} + \bm{\delta}_{l'} \uparrow}^{\phantom{\dag}}}$. Eventually, we have the effective pairing interaction $\bar{P}_\alpha = P_\alpha - \widetilde{P}_\alpha$ as well as the effective zero-frequency pair-pair correlation function $\bar{\mathcal{D}}_\alpha (\mathbf{i},\mathbf{j})=\mathcal{D}_\alpha(\mathbf{i},\mathbf{j}) -\widetilde{\mathcal{D}}_\alpha (\mathbf{i},\mathbf{j})$.
The appearance of negative effective pairing interaction may indicate that the pairing symmetry is suppressed by other competing states.

We also define the effective \textit{zero-frequency} pair-pair structure factor for DQMC,
\begin{equation}
\bar{\mathcal{D}}_\alpha (\mathbf{q})=\frac{1}{N_{s}} \sum_{\mathbf{i}, \mathbf{j}}
e^{i \mathbf{q} \cdot (\mathbf{i}-\mathbf{j})} \bar{\mathcal{D}}_\alpha (\mathbf{i}, \mathbf{j})\ .
\label{shi3}
\end{equation}
In particular, we use $P^\text{PDW}_d (\mathbf{q}) \equiv \bar{\mathcal{D}}_d (\mathbf{q})$ to understand the effects of the charge-density modulation on the $d$-wave pair-density-wave (PDW) order.
In the simulations, when the peak of $P_d^\text{PDW}(\mathbf{q})$ is located at zero momentum, it
indicates a lack of PDW state in the system. Otherwise, there may be a PDW
state \cite{annurev050711,Huang2022}.

As magnetic excitation possibly plays an important role for the superconductivity mechanism in strong electron correlation systems, we also study the spin susceptibility in the $z$ direction at zero frequency in the $\mathcal{P}=3$ model,
\begin{eqnarray}
\chi_{s}(\mathbf{q})=\frac{1}{N_s}\int_0^\beta d\tau \sum_{\mathbf{i},\mathbf{j}}
e^{i\mathbf{q}\cdot(\mathbf{i}-\mathbf{j})} \langle \textrm{m}_{\mathbf{i}}(\tau) \
\textrm{m}_{\mathbf{j}}(0)\rangle_{T}\ ,
\end{eqnarray}
where $m_{\mathbf{i}}(\tau)$=$e^{H\tau}m_{\mathbf{i}}(0)e^{-H\tau}$ with
$m_{\mathbf{i}}(0)=c_{\mathbf{i} \uparrow}^\dag c_{\mathbf{i} \uparrow}^{\phantom{\dag}} - c_{\mathbf{i} \downarrow}^\dag c_{\mathbf{i} \downarrow}^{\phantom{\dag}}$.

{\it{DMRG method:}} At zero temperature, we employ the DMRG method to investigate the model Hamiltonian on a cylinder with $8,192$ SU($2$) bases at most, equivalent to about $25,000$ U($1$) bases, and guarantee that the truncation error is less than $10^{-5}$.
We also examine the pairing-symmetry transition directly by investigating the \textit{static} pair-pair structure factor
\begin{equation}
\mathcal{S}_\alpha(\mathbf{q})=\frac{1}{N_s} \sum_{\mathbf{i},\mathbf{j}} e^{i\mathbf{q} \cdot(\mathbf{i}-\mathbf{j})} \left\langle \Delta_\alpha^\dag (\mathbf{i}, 0)\Delta_\alpha^{\phantom{\dag}} (\mathbf{j}, 0) \right\rangle,
\label{shi3s}
\end{equation}
where the statistic average at a finite temperature and zero frequency in Eq.~\eqref{shi3} is replaced with the ground-state expectation value at zero temperature here. Similarly, we also calculate the uncorrelated single-particle contribution $\widetilde{\mathcal{S}}_\alpha (\mathbf{q})$ and define the effective static pair-pair structure factor as $\bar{\mathcal{S}}_\alpha (\mathbf{q}) = \mathcal{S}_\alpha (\mathbf{q}) - \widetilde{\mathcal{S}}_\alpha (\mathbf{q})$.
In the calculation, we target the lowest-energy zero-magnetic-momentum state with a specified even number of electrons. Thus, the number of electrons for any species is also preserved and the spin fluctuations remain negligible. 
In this work, we use the effective zero-momentum pair-pair structure factors $\bar{S}_s \equiv \bar{\mathcal{S}}_s (\mathbf{q} = (0, 0))$ and $\bar{S}_d \equiv \bar{\mathcal{S}}_d (\mathbf{q} = (0, 0))$, and the emerging peak of $\bar{\mathcal{S}}_d (\mathbf{q})$ at a finite momentum $\mathbf{q} \ne (0, 0)$ to identify the $s$ and $d$-wave pairing as well as the $d$-wave PDW, respectively.

To clearly illustrate how the pairing-symmetry transition happens at zero temperature, we further decompose Cooper pair modes from the two-particle density matrix, defined as~\cite{PhysRevLett.129.177001}
\begin{eqnarray}
\rho (\mathbf{i} \bm{\delta}_l, \mathbf{j} \bm{\delta}_{l'}) = \left \langle{\mathcal{C}_{\mathbf{i} \bm{\delta}_l}^\dag \mathcal{C}_{\mathbf{j} \bm{\delta}_{l'}}^{\phantom{\dag}}}\right\rangle\ ,
\end{eqnarray}
where $\mathcal{C}_{\mathbf{i} \bm{\delta}_l}$ is consistent with the definition in Eq.~\ref{shi22}. We exclude the overlapping parts for either $\mathbf{i} = \mathbf{j}$, or $\mathbf{i} = \mathbf{j} + \bm{\delta}_{l'}$, or $\mathbf{j} = \mathbf{i} + \bm{\delta}_l$, giving rise to the local contributions from density and spin correlations.
Since $\rho$ is Hermitian, it can be diagonalized with real eigenvalues $\lambda_n$, that is,
\begin{eqnarray}
\rho (\mathbf{i} \bm{\delta}_l, \mathbf{j} \bm{\delta}_{l'}) = \sum_n \lambda_n \zeta^*_n(\mathbf{i} \bm{\delta}_l) \zeta_n(\mathbf{j} \bm{\delta}_{l'})\ .
\end{eqnarray}
The eigenvector $\zeta_n (\mathbf{i} \bm{\delta}_l)$ are referred to as macroscopic wave functions of Cooper pair modes.
The dominant mode with the largest eigenvalue is labeled by $n=0$.

{\it{CPQMC method:}}
To further demonstrate that the system may exhibit long-range superconducting correlations for the $s$ wave pairing, we also check the long-range part of the ground-state pair-correlation function using the CPQMC method~\cite{HUANG2019310, PhysRevLett.74.3652}. The CPQMC method has been successfully used to calculate the ground-state energy and other observables in various systems~\cite{HUANG2019310,PhysRevLett.74.3652}.
We investigate the long-range superconducting correlations of dominant $s$-wave pairing symmetry by defining the pair-pair correlation function at zero temperature, which is written as
\begin{equation}
C_\alpha(\mathbf{r})=\frac{1}{NN_r} \sum_{\mathbf{i},\mathbf{j}} \sum_{|\mathbf{j}-\mathbf{i}|=r}  \left\langle \Delta_\alpha^\dag (\mathbf{i}, 0)\Delta_\alpha^{\phantom{\dag}} (\mathbf{j}, 0) \right\rangle,
\label{shi3s}
\end{equation}
Here, $r$ is the distance between site $\mathbf{i}$ and site $\mathbf{j}$. The $N_r$ is
the total number of distance $r$.
Similarly, we also define the uncorrelated single-particle contribution $\widetilde{C}_\alpha (\mathbf{r})$ and discuss the vertex contributions $\bar{C}_\alpha (\mathbf{r}) = {C}_\alpha (\mathbf{r}) - \widetilde{C}_\alpha (\mathbf{r})$.

{\textbf{Data availability}}

Data are available from the authors upon reasonable request.

{\textbf{Code availability}}

DQMC and DMRG codes used for the data processing and other findings of this study are available upon request.

{\textbf{Author contributions}}

B.H. convinced the projects. C.C., B.H., S.H., and T.M. directed the project. C.C., R.M., T.M., and H.Q.L. developed the DQMC and CPQMC codes and performed the simulations. P. Z. and S.H. developed the DMRG code and performed the simulations. C.C., S.H., and B.H. prepared the manuscript. All authors discussed the results and contributed to the manuscript.

{\textbf{Competing interests}}

The authors declare no competing interests.

{\textbf{Acknowledgements}} We thank Rubem Mondaini and Xuefeng Zhang for useful discussions.
This work was supported by National Natural Science Foundation of China (Grants No. 12088101 and No. 11974049) and  the NSAF (Grant No. U2230402).
The numerical simulations in this work were performed at the HSCC of Beijing Normal University and Tianhe2-JK in Beijing Computational Science Research Center.

\bibliography{reference}



\newpage

\newpage
\onecolumngrid
\vspace{3cm}

\begin{center}
{\bf\large Supplemental Material for \\``Charge Stripe Manipulation of Superconducting Pairing Symmetry Transition''}
\end{center}
\vspace{0.5cm}

\emergencystretch=\maxdimen
\hyphenpenalty=10000
\hbadness=10000
\bibliographystyle{naturemag}  
\renewcommand*{\thefigure}{S\arabic{figure}}

\title{Supplemental Material for \\``Charge Stripe Manipulation of Superconducting Pairing Symmetry Transition''}
\author{Chao Chen}
\affiliation{Department of Physics, Beijing Normal University, Beijing 100875, China\\}
\affiliation{Beijing Computational Science Research Center, Beijing 100084, China\\}
\author{Peigeng Zhong}
\affiliation{Beijing Computational Science Research Center, Beijing 100084, China\\}
\author{Xuelei Sui}
\affiliation{Beijing Computational Science Research Center, Beijing 100084, China\\}
\author{Runyu Ma}
\affiliation{Department of Physics, Beijing Normal University, Beijing 100875, China\\}
\author{Ying Liang}
\affiliation{Department of Physics, Beijing Normal University, Beijing 100875, China\\}
\affiliation{Key Laboratory of Multiscale Spin Physics, Ministry of Education, Beijing 100875, China\\}
\author{Shijie Hu}
\affiliation{Beijing Computational Science Research Center, Beijing 100084, China\\}
\affiliation{Department of Physics, Beijing Normal University, Beijing 100875, China\\}
\author{Tianxing Ma}
\affiliation{Department of Physics, Beijing Normal University, Beijing 100875, China\\}
\affiliation{Key Laboratory of Multiscale Spin Physics, Ministry of Education, Beijing 100875, China\\}
\author{Hai-Qing Lin}
\affiliation{Center for Correlated Matter and School of Physics, Zhejiang University, Hangzhou 310058, China\\}
\affiliation{Beijing Computational Science Research Center, Beijing 100084, China\\}
\affiliation{Department of Physics, Beijing Normal University, Beijing 100875, China\\}
\author{Bing Huang}
\affiliation{Beijing Computational Science Research Center, Beijing 100084, China\\}
\affiliation{Department of Physics, Beijing Normal University, Beijing 100875, China\\}

\maketitle

\begin{center}
  \textbf{1. Superconducting pairing interaction at \tck{charge-stripe} period \tck{$\mathcal{P}=2$, $4$} }
\end{center}

\begin{figure}[h]
\includegraphics[scale=0.3,trim = 50 50 0 40, clip]{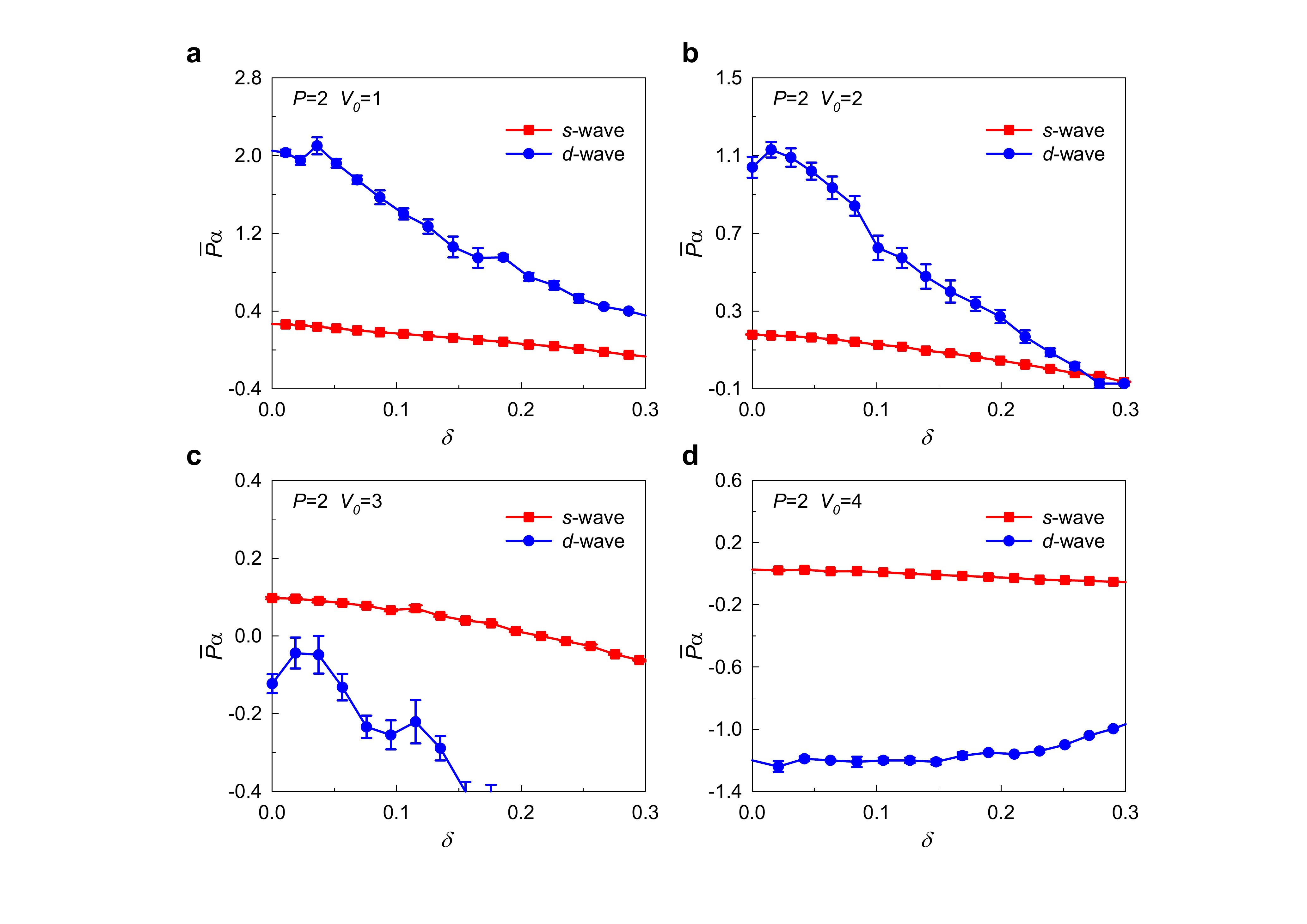}
\caption{(Color online) DQMC-calculated effective pairing interaction $\bar{P}_{\alpha}$ as a function of hole-doping \tck{concentration} $\delta$ at $T=t/5$ and $U/t=4$ with \tck{a charge-stripe} period $\mathcal{P}=2$ on a $L=12$ lattice for the different \tck{charge-stripe amplitudes of (a) $V_0=1$, (b) $V_0=2$, (c) $V_0=3$, and (d) $V_0=4$}.
}
\label{Fig:stripe2}
\end{figure}

\begin{figure}[h]
\includegraphics[scale=0.3,trim = 50 50 0 40, clip]{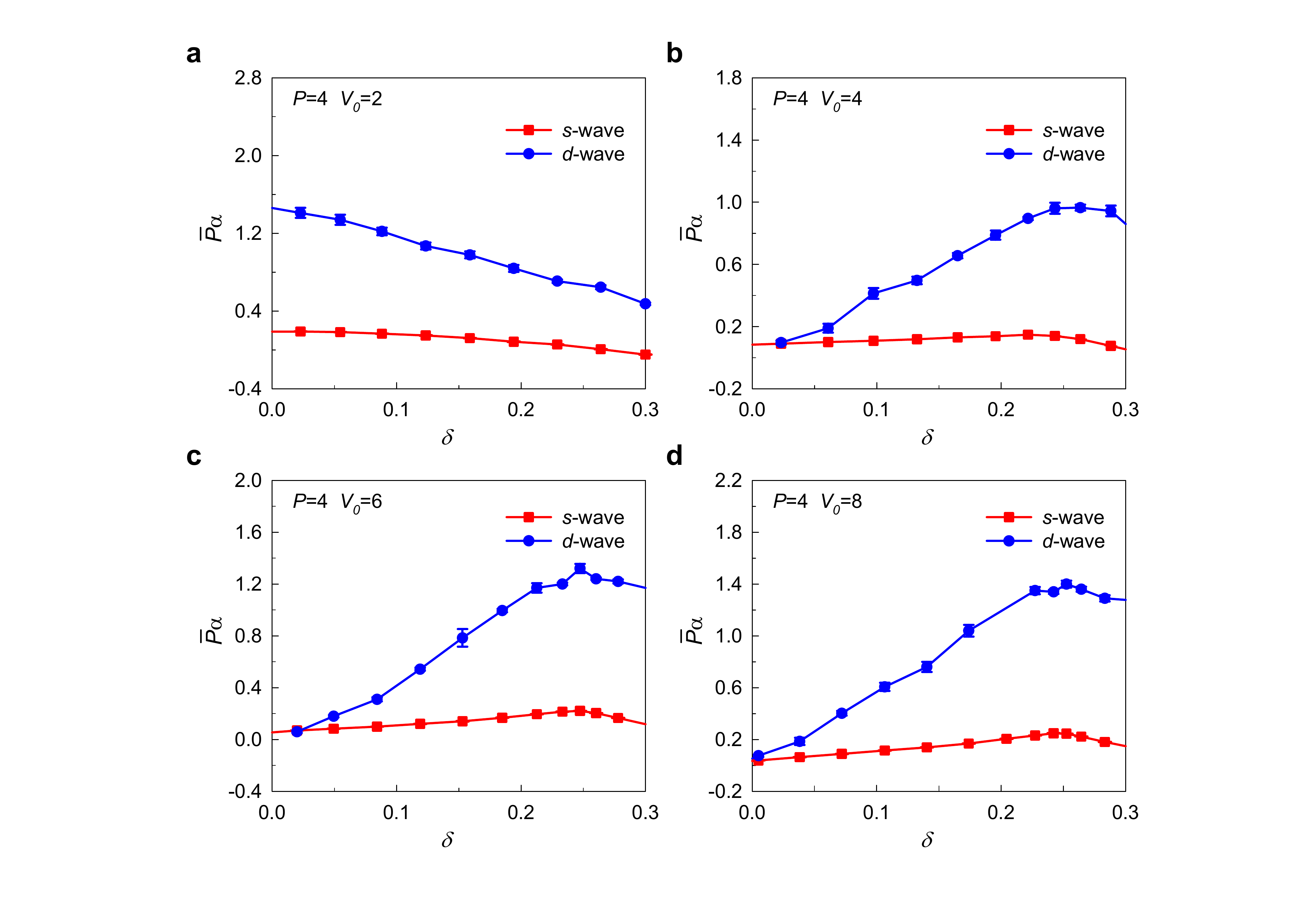}
\caption{(Color online) DQMC-calculated effective pairing interaction $\bar{P}_{\alpha}$ as a function of hole-doping \tck{concentration} $\delta$ at $T=t/5$ and $U/t=4$ with \tck{a charge-stripe} period $\mathcal{P}=4$ on a $L=12$ lattice for the different \tck{charge-stripe amplitudes of (a) $V_0=2$, (b) $V_0=4$, (c) $V_0=6$, and (d) $V_0=8$}.
}
\label{Fig:stripe4}
\end{figure}

\tck{To} systematically study the impact of different stripe \tck{periods}, we also calculate the superconducting pairing interaction at $\mathcal{P}=2$, \tck{$4$}. Interestingly, as the \tck{change-stripe amplitude} is enhanced from \tck{$V_0=2$ to $3$} in Fig.~\ref{Fig:stripe2} for $\mathcal{P}=2$, the \tck{$s$-wave} pairing symmetry is dominant for all the hole-doping concentrations. Thus, the \tck{dominant pairing-symmetry transition} is \tck{more likely} to occur at period $\mathcal{P}=2$ than \tck{at} $\mathcal{P}=3$, despite \tck{a} lack of observed $\mathcal{P}=2$ state in the current experiments.

In Fig.~\ref{Fig:stripe4}, we \tck{have} studied the hole-doping dependence of the effective pairing interaction $\bar{P}_\alpha$ at $T=t/5$ and $U/t=4$ with different $V_0$ for $\mathcal{P}=4$.
As shown in Fig.~\ref{Fig:stripe4}, the \tck{$d$-wave} pairing symmetry is always robust for different $V_0$ \tck{in} hole-doping regions. This result is consistent with the dominant $d$ pairing symmetry of the cuprates in the experiments.
Overall, \tck{our} conclusion is consistent with the results
of \tck{the previous} DCA study by Maier et al \cite{PhysRevLett.104.247001}. Their paper suggests that charge inhomogeneity ($V_0=0\sim0.6$) \tck{does not affect} the pairing correlations when $\mathcal{P} = 4$ and $\delta=0.125$. In addition, in our simulations, we \tck{can also see} that $\bar{P}_d$ drops from \tck{about $1.2$ to $0.6$ when} going from $V_0 = 2$ to $8$ at $\delta=0.125$.

\begin{center}
  \textbf{2. Four-band model for nickelates}
\end{center}

To confirm the robustness of \tck{the conclusion in the minimal} single-band model, we select the
representative four-band model and parameters in \tck{Ref.~\cite{GuGu2020}}. The model includes \tck{$d_{x^2-y^2}$} and $R$ \tck{$d_{xy}$/$d_{3z^2-r^2}$} orbitals and \tck{interstitial $s$} (i-s) orbitals. Such a model could contain both out-of-plane rare-earth orbitals and oxygen orbital contributions. The Hamiltonian can be written as,

\begin{eqnarray}
\label{Hamiltonian}
H=\sum_{i, a, mm' , \sigma} H_0( a)_{mm'} c_{ i m \sigma}^\dagger c_{(i+a) m' \sigma}^{\phantom{\dagger}}
+U_{Ni} \sum_{{\bf i1}}   n_{{\bf i1}\uparrow}  n_{{\bf i1}\downarrow} \nonumber +V_0\sum_{\text{mod}(i_y,3)=0} (n_{{\bf i1}\uparrow}+n_{{\bf i1}\downarrow})+H_{\mu} .
\end{eqnarray}

\begin{eqnarray}
\label{Hamiltonian}
H_{\mu}={\mu}_1 \sum_{\mathbf{i2}\sigma} n_{\mathbf{i2}\sigma} +{\mu}_2 \sum_{\mathbf{i3}\sigma} n_{\mathbf{i3}\sigma} +{\mu}_3 \sum_{\mathbf{i4}\sigma} n_{\mathbf{i4}\sigma} - {\mu} \sum_{\mathbf{im}\sigma} n_{\mathbf{im}\sigma}  ,
\end{eqnarray}

\begin{equation}
H_0(a)={
\begin{array}{cc}
& \begin{array}{cccc}d_{x^2-y^2} & i-s & d_{xy} & d_{3z^2-r^2}\end{array}\\
\begin{array}{c} d_{x^2-y^2}\\i-s\\d_{xy}\\d_{3z^2-r^2}\end{array}&
\left(\begin{array}{cccc}
   -0.37 &  -0.22  &  0.03  & -0.02 \\
   -0.22 &  -0.24  &  0.68  &  0.45 \\
    0.03 &   0.68  & -0.08  &  0    \\
   -0.02 &   0.45  &  0     & -0.19  \end{array}\right)
\end{array}
}
\end{equation}

Here, $V_0$ is an additional site potential exerted on a set of rows for \tck{the $d_{x^2-y^2}$ band}. ${\mu}_1$, ${\mu}_2$ and ${\mu}_3$ represent the on-site energy difference of the interstitial $s$ orbital, \tck{$d_{xy}$/$d_{3z^2-r^2}$} orbitals \tck{with} respect to \tck{the}-$d_{x^2-y^2}$ orbital, respectively.
$m=1$, $2$, $3$, $4$ and $m'=1$, $2$, $3$, $4$ labels different orbitals, $a$ denotes nearest-neighbor sites, $\sigma$ labels spins. $n_{\mathbf{im}\sigma}=c_{\mathbf{im}\sigma}^{\dagger}c_{\mathbf{im}\sigma}$ is the \tck{particle number operator for electrons with} spin $\sigma$ at $m$ orbital and the coulomb repulsion $U_\text{Ni}$ is solely applied on the \tck{Ni}-$d_{x^2-y^2}$ orbital.

\begin{figure}[tbp]
\includegraphics[scale=0.4,trim = 140 50 60 25, clip]{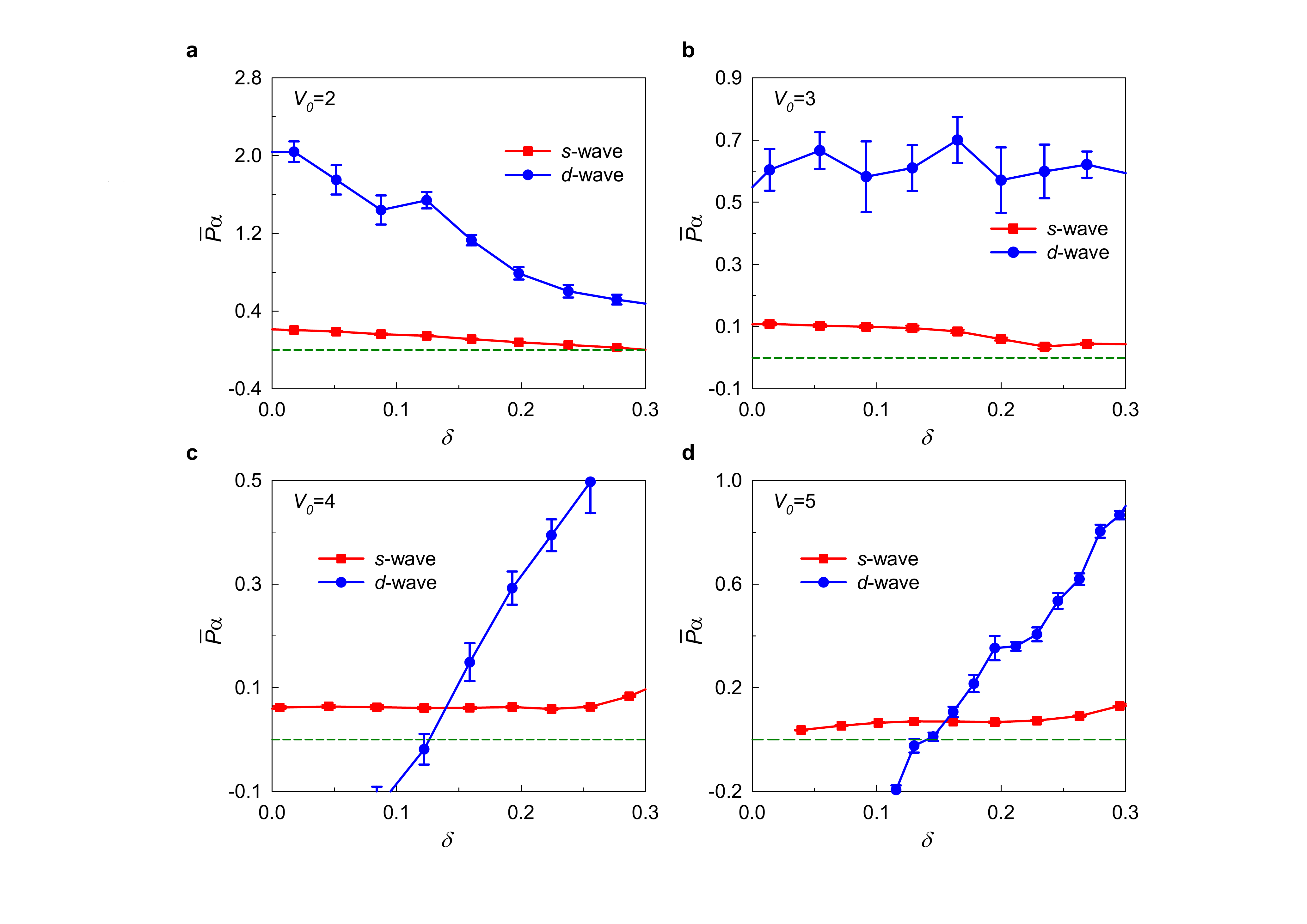}
\caption{DQMC-calculated effective pairing interaction $\bar{P}_{\alpha}$ as a function of hole-doping concentration $\delta$ at $T=t/5$ and $U/t=4$ with stripe period $\mathcal{P}=3$ on a $N=4\times L^2=144$ lattice at the different stripe potential (a) $V_0$=2, (b) $V_0$=3, (c) $V_0$=4, and (d) $V_0$=5, for the four-band model.
}
\label{Fig:4band}
\end{figure}

In Figs.~\ref{Fig:4band}, we calculate the $\delta$-dependence of the
effective pairing interaction $\bar{P}_{\alpha}$ at different $V_0$ under the four-band Hubbard model. Similar to the results of \tck{the} single-band model, we can also observe the $d$-$s$ wave transition. When $V_0=0 \sim 3$ in \tck{Figs.~\ref{Fig:4band}(a)-(b)}, the $\bar{P}_{d}$ is always dominant under all hole-doping concentrations. However, when $V_0$ increases to $4$ in \tck{Figs.~\ref{Fig:4band}(c)}, the $\bar{P}_{s}$ is more stable than $\bar{P}_{d}$ in the moderate $\delta$ ($\tck{0}< \delta \leq0.15$).
From our calculations, we can conclude that the $d$-$s$ wave transition
may not depend on the model itself (single or multiple bands), but rely on the intrinsic interplay between
\tck{the} hole-doping concentration $\delta$, charge-stripe period $\mathcal{P}$ and
charge-stripe amplitude $V_0$, which might partially account for some experimental observations \cite{Gu2020,Wang2021,arxiv.2201.12971,arXiv:2201.10038}.

\begin{figure}[h]
\includegraphics[scale=0.44,trim = 100 260 100 180, clip]{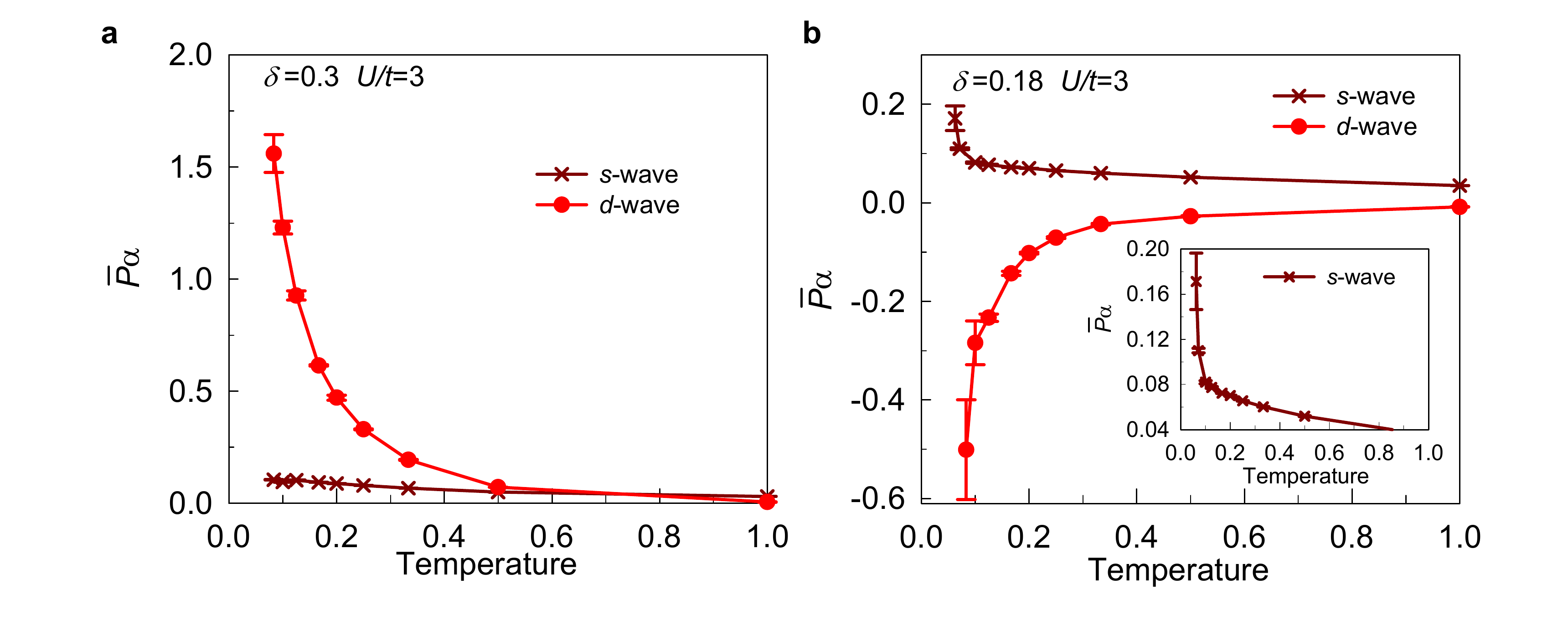}
\caption{DQMC-calculated effective pairing interaction $\bar{P}_{\alpha}$ as a
function of temperature for different pairing symmetry at $U/t$=3, $V_0$=6 and
$L=12$, with (a) $\delta=0.3$, (b) $\delta=0.18$. Inset of (b): $\bar{P}_{\alpha}$
of $s$ pairing symmetry versus temperature.
}
\label{Fig:low-DQMC}
\end{figure}

\begin{center}
  \textbf{3. Superconducting pairing interaction at a lower temperature}
\end{center}

The effective pairing interaction $\bar{P}_{\alpha}$ at lower temperature has been calculated using DQMC on a torus of $12\times12$ lattice for $U/t$=3, reaching $T=t/12$. In Fig.~\ref{Fig:low-DQMC} (a), when $\delta=0.3$ and $V_0$=6, we can see that \tck{the} dominant $\bar{P}_{\alpha}$ for the normal $d$-wave rises more rapidly with the decrease of temperature than that for the \tck{$s$ wave}. However, when $\delta=0.18$ and $V_0$=6, the trend is reversed. As shown in \tck{Fig.~\ref{Fig:low-DQMC}(b)}, \tck{$\bar{P}_s$} for the \tck{$s$ wave} slowly increases with decreasing temperature, while \tck{$\bar{P}_d$} for the \tck{$d$ wave} quickly decreases as the temperature is lowered, reflecting the $d$ pairing symmetry suppressed by other competing pairing channels or phases. This \tck{demonstrates} that the $s$ pairing symmetry \tck{can indeed} be dominant even at \tck{lower temperatures}. Besides, our calculations indicate that this dominant \tck{$s$-wave} pairing symmetry can be further enhanced by larger coulomb repulsion $U$ (\tck{Fig.~4} in the main text). So, the dominant \tck{pairing interaction} for the \tck{$s$-wave} pairing symmetry could diverge in the \tck{thermodynamic} limit at sufficiently low temperatures.

\begin{figure}[tbp]
\includegraphics[scale=0.4,trim = 50 100 30 50, clip]{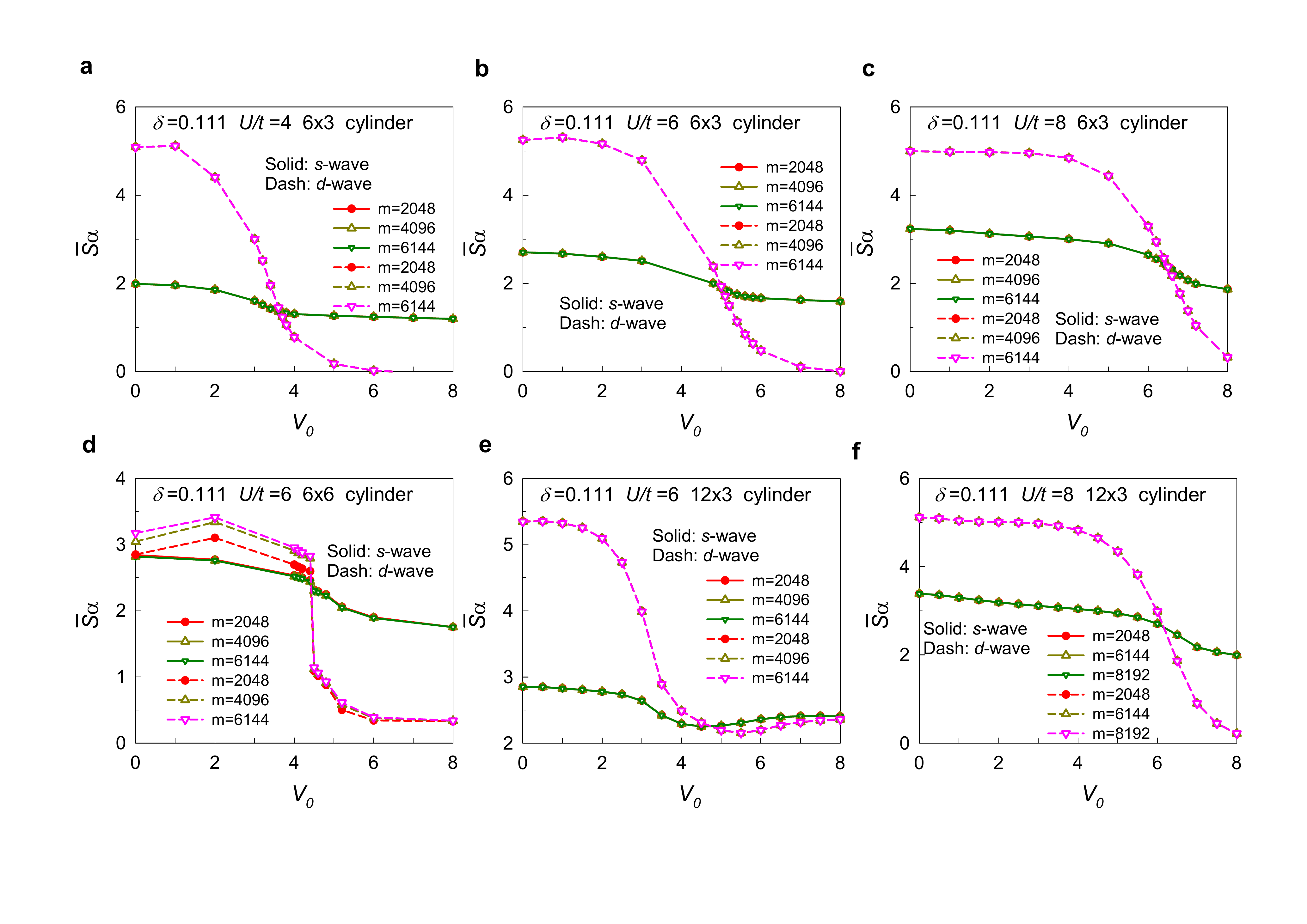}
\caption{DMRG-calculated effective \tck{zero-momentum} pair-pair \tck{structure factor $\bar{S}_\alpha$} as
a function of $V_0$ on the $L_x \times L_y$ cylinder. \tck{With the hole-doping concentration $\delta=1/9\approx0.111$ and period $\mathcal{P}=3$, we show the results for six typical parameter sets}: (a)~$U/t=4$, (b)~$6$ and (c)~$8$ on the $6\times3$ cylinder, (d)~$U/t=6$ on the $6\times6$ cylinder, (e)~$U/t=6$ and (f)~$8$ on the $12\times3$ cylinder. For $L_y=3$ in (a)-(c) and (e)-(f), the data collapse \tck{demonstrates} that the bond dimension $m \le 6144$ is large enough to guarantee the numerical precision. For \tck{$L_y = 6$}, a small discrepancy can be observed, however, it does not affect our conclusion \tck{within error bars}.
}
\label{Fig:zero-DMRG}
\end{figure}

\begin{center}
  \textbf{4.~DMRG-calculated effective zero-momentum pair-pair structure factor}
\end{center}

In Fig.~\ref{Fig:zero-DMRG}, we exhibit the transition from the $d$-wave pairing to the extended $s$-wave pairing for the typical cases of different cylinder sizes and interaction strengths. From \tck{Figs.~\ref{Fig:zero-DMRG}(a)-(c)} for the $6 \times 3$ cylinder, we can \tck{see} that the critical transition point $V_{0,c}$, the place where the curves of the two effective zero-momentum pair-pair structure factors $\bar{S}_d(0)$ and $\bar{S}_{s}(0)$ intersect, becomes larger with increasing $U$ from $4$ to $8$. However, with the fixed $U$ and $L_x$, the position of the $V_{0,c}$ depends less on the growth of $L_y$, as shown in Fig.~\ref{Fig:zero-DMRG}(d) and Fig. 2(e) of the \tck{main} text. If the aspect ratio $L_x / L_y$ is too large, e.g., a long and thin cylinder with $L_x=12$ and $L_y=3$ in Figs.~\ref{Fig:zero-DMRG}(e)-(f), it is very difficult to determine the physics on the large-$V_0$ side, since two structure factors become comparable in quasi-one dimension, although the pairing-symmetry transition can also be observed. Moreover, from the data collapse of numerical results for different bond dimensions $m$ in the SU($2$) DMRG, we conclude that $m \le 6144$ is large enough for $L_y \le 6$, beyond which DQMC could provide an excellent complement.

\begin{figure}[h]
\includegraphics[scale=0.6,trim = 160 200 200 50, clip]{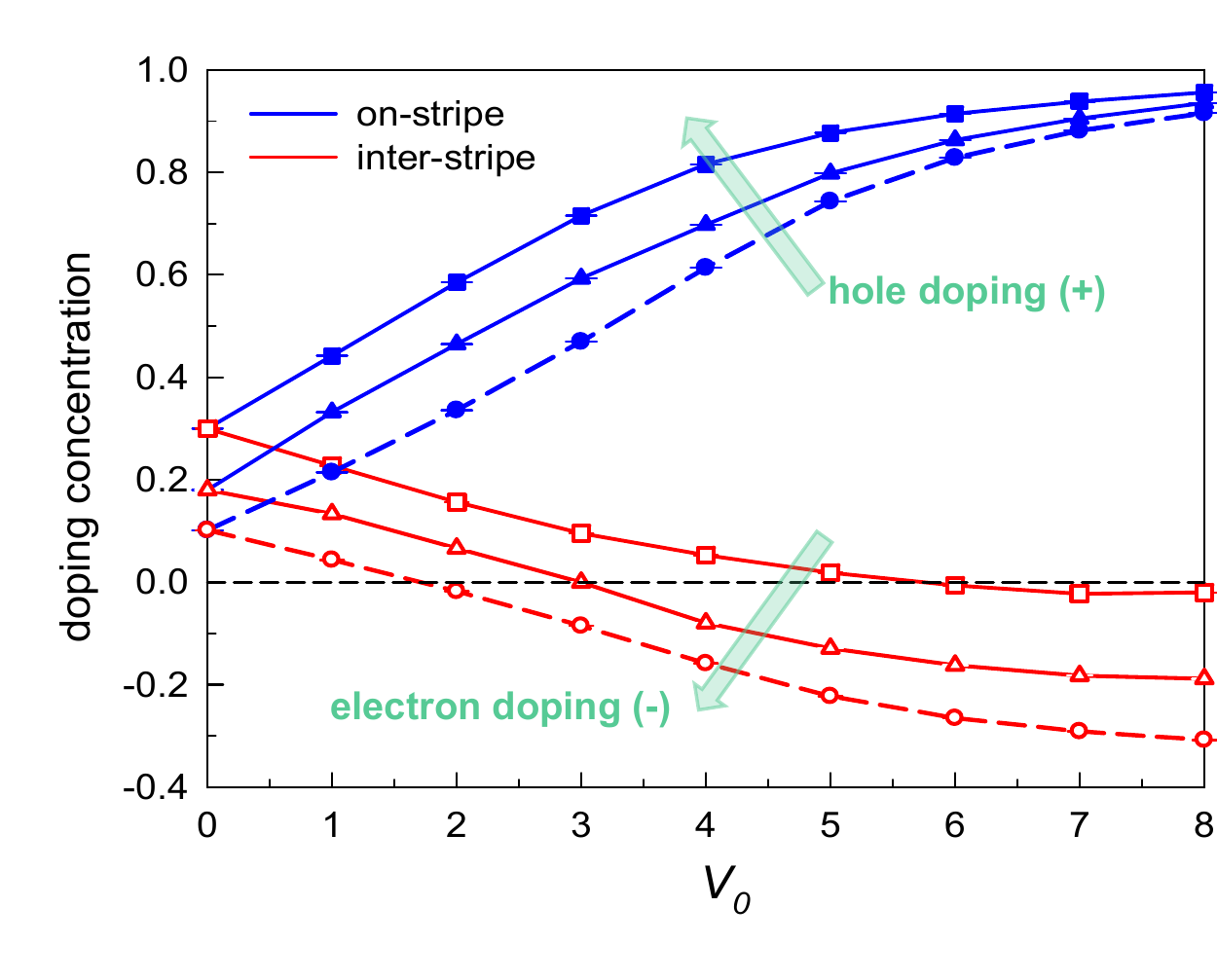}
\caption{(a) DQMC-calculated doping distribution at the on-stripe region [blue sites in Fig. 1 (a)] and the inter-stripe region [red sites in Fig. 1 (a)]. Total doping concentration of the system is fixed at $\delta$=0.3 (square-symbol), 0.2 (triangle-symbol) and 0.1 (circular-symbol). Parameters are set to $T=t/5$, $U/t=4$, $L=12$ and $\mathcal{P}=3$. Note that $\delta>0$ ($\delta<0$) means the inter-stripe region is hole (electron) doped.
}
\label{Fig:density}
\end{figure}

\begin{center}
  \textbf{5. DQMC-calculated doping distribution}
\end{center}

As shown in Fig.~\ref{Fig:density}, we have plotted the hole-doping redistribution
in the on-stripe [blue sites, Fig.~1(a) in the main text] and inter-stripe [red sites, Fig.~1(a) in the main text]
regions as a function of $V_0$, in which the total hole-doping concentration of
the entire lattice is fixed at \tck{$\delta=0.3$, $0.2$ and $0.1$}. When $V_0=0$, the system
is homogeneous, and the $\delta$(on-stripe) and $\delta$(inter-stripe) are equal.
As $V_0$ increases, the charge redistribution occurs between the on-stripe and
inter-stripe regions, resulting in the different $\delta$(on-stripe) and
$\delta$(inter-stripe). Ultimately, for $V_0 \ge 8$, the on-stripe region is nearly
empty and the inter-stripe region is saturated. Interestingly, for $V_0>5$ at
\tck{$\delta=0.3$}, the inter-stripe region is nearly half-filled ($\delta$(inter-stripe)
$\approx$ 0), creating \tck{the} strongest correlation effect.

\begin{center}
  \textbf{6. $d$-wave pair-density-wave state}
\end{center}

\begin{figure}[tbp]
\includegraphics[scale=0.65,trim = 0 10 0 10, clip]{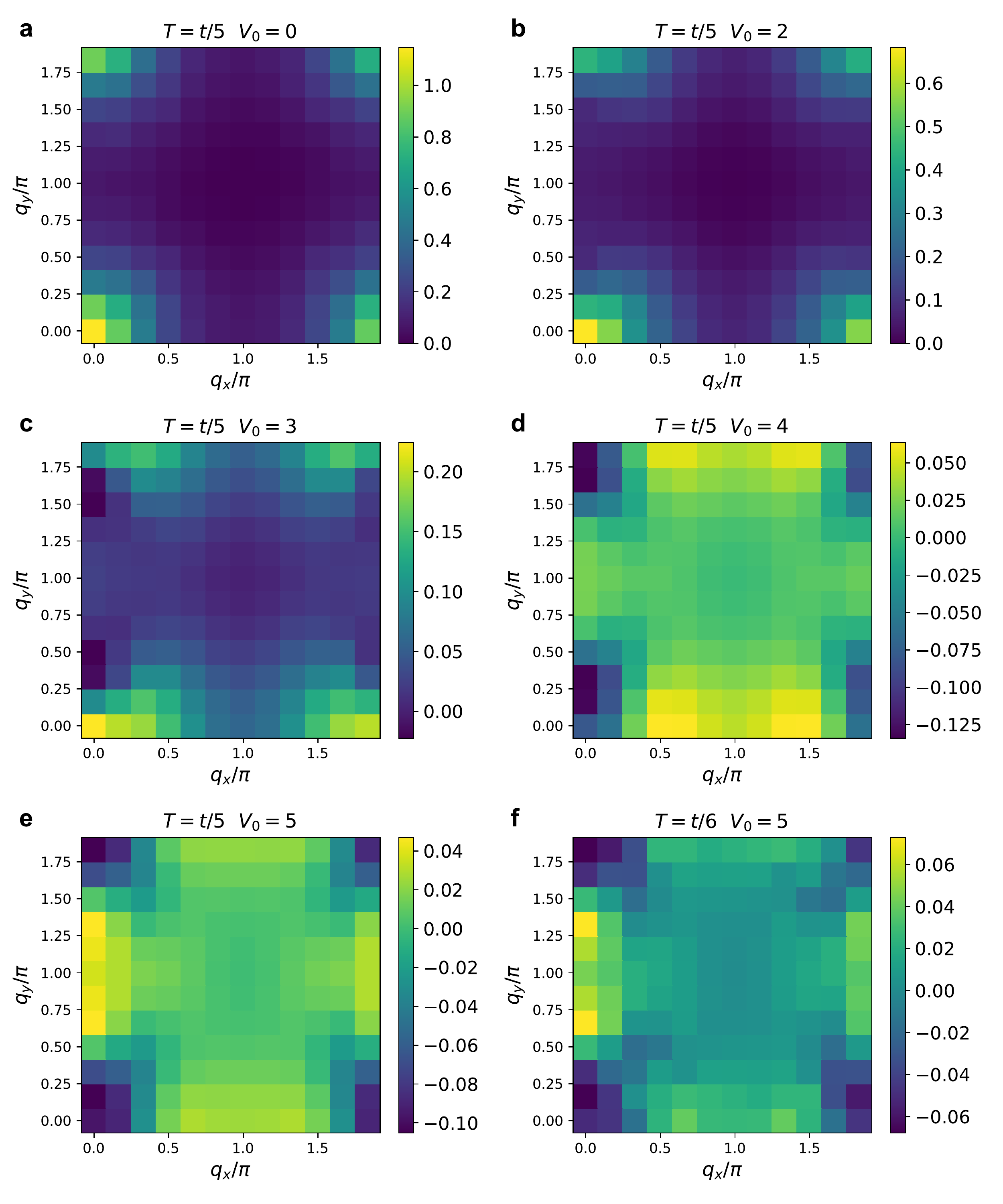}
\caption{DQMC-calculated \tck{zero-frequency pair-pair} structure factor of $d$-wave PDW \tck{$P_{d}^\text{PDW}(\mathbf{q})$} in the first Brillouin zone at $\delta=0.18$, $U/t=4$ with (a) $V_0=0$ and $T=t/5$,
(b) $V_0=2$ and $T=t/5$, (c) $V_0=3$ and $T=t/5$, (d) $V_0=4$ and $T=t/5$,
(e) $V_0=5$ and $T=t/5$, (f) $V_0=5$ and $T=t/6$.
}
\label{Fig:PDW-DQMC}
\end{figure}

Here, we \tck{choose} the typical cases of $\delta=0.18$, because this doping concentration
can realize \tck{the} $d$-$s$ wave transition by manipulating $V_0$.
In \tck{Figs.~\ref{Fig:PDW-DQMC}(a)-(c)} for $V_0=0\sim 3$, we can \tck{see} that
the $P_{d}^\text{PDW} (\mathbf{q})$ is largest at $\mathbf{q}=(0,0)$, indicating no $d$-wave PDW state
in the $d$-wave dominated region. When $V_0$ is larger than \tck{$3$ in Figs.~\ref{Fig:PDW-DQMC}(d)-(f)},
the peaks of $P_{d}^\text{PDW}(\mathbf{q})$ \tck{move away} from zero \tck{momentum} and the system shows a tendency to form \tck{the} PDW state.
However, the $\bar{P}_{s}$ is always \tck{larger} than the peak value of $P_{d}^\text{PDW}(\mathbf{q})$. This demonstrates that the $s$ wave is robust in the phase diagram despite the existence of competing $d$-wave PDW.
Here, we \tck{only} list the typical hole-doping concentration $\delta=0.18$.
Other hole-doping concentrations also \tck{give} similar results that
competing $d$-wave PDW state \tck{can} form at low temperatures, but the \tck{$s$ wave}
is still robust in the phase diagram.

\begin{figure}[tbp]
\includegraphics[scale=0.65,trim = 0 10 0 10, clip]{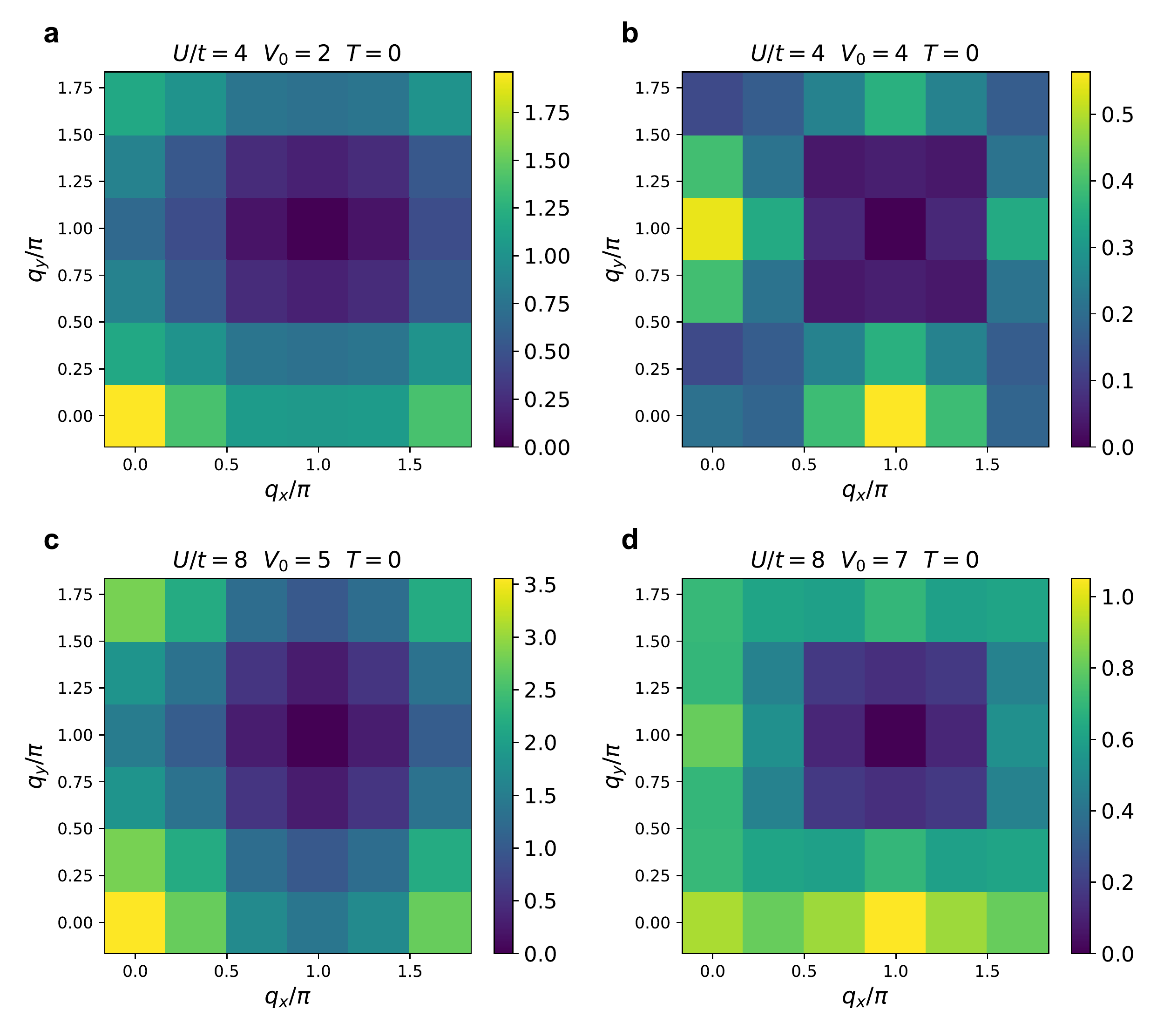}
\caption{DMRG-calculated effective $d$-wave \tck{static} pair-pair structure factor $\bar{\mathcal{S}}_d(\mathbf{q})$ in the first Brillouin zone for the $6\times6$ cylinder at zero temperature $T=0$, with the \tck{hole-doping concentration} $\delta=1/9\approx0.111$. We choose (a) $V_0=2$ and $U/t=4$, (b) $V_0=4$ and $U/t=4$, (c) $V_0=5$ and $U/t=8$, (d) $V_0=7$ and $U/t=8$. In (a) and (c), the maximum values of $\bar{\mathcal{S}}_d(\mathbf{q})$ are situated at $\mathbf{q}=(q_x,\ q_y)=(0,\ 0)$, corresponding to the normal $d$-wave pairing. In (b) and (d), the maximum values appear at $\mathbf{q}=(\pi,\ 0)$, indicating an exotic $d$-wave PDW. The DMRG bond dimension $m=8192$.
}
\label{Fig:PDW-DMRG}
\end{figure}

At zero temperature and typical doping $\delta=1/9\approx 0.111$, we calculate the effective $d$-wave \tck{static} pair-pair structure factor $\bar{S}_d(\mathbf{q})$ in the first Brillouin zone (FBZ), using DMRG method, \tck{targeting the ground state with an equal number of electrons for each species}. First, along the cutting line $U/t=4$ as shown in Fig.~\ref{Fig:PDW-DMRG}(a), for small $V_0=2$, $\bar{S}_d (\mathbf{q})$ reaches the maximum at zero momentum $\mathbf{q}=(q_x,\ q_y)=(0,\ 0)$, indicating the normal $d$-wave pairing.
For larger $V_0=4$ in Fig.~\ref{Fig:PDW-DMRG}(b), the maximum of $\bar{S}_d(\mathbf{q})$ is situated at a finite momentum $\mathbf{q}=(\pi,\ 0)$, suggesting a possible $d$-wave PDW in the ground state. However, the maximum value $\bar{S}_d(\mathbf{q}=(\pi,\ 0)) \approx 0.5633$ is \tck{about} $1/3$ of $\bar{S}_{s}(\mathbf{q}=0) \approx 1.4639$, so the ground state energetically favors the $s$-wave pairing, keeping consistent with Fig.~3\tck{(f)} in the main text.
Similarly, along another cutting line of larger $U/t=8$, we also observe the transition from the normal $d$-wave pairing to the extended $s$-wave pairing \tck{as} $V_0$ grows from $5$ to $7$. In this case, we find the relatively weak signal of the $d$-wave PDW as well, in which the maximum of $\bar{S}_d(\mathbf{q})$ is located at the momentum $\mathbf{q}=(\pi,\ 0)$, as shown in Fig.~\ref{Fig:PDW-DMRG}(d).

\begin{center}
  \textbf{7. The calculations about long-range $s$-wave superconducting correlation using CPQMC and DMRG}
\end{center}

\begin{figure}[tbp]
\includegraphics[scale=0.42,trim = 40 120 30 100, clip]{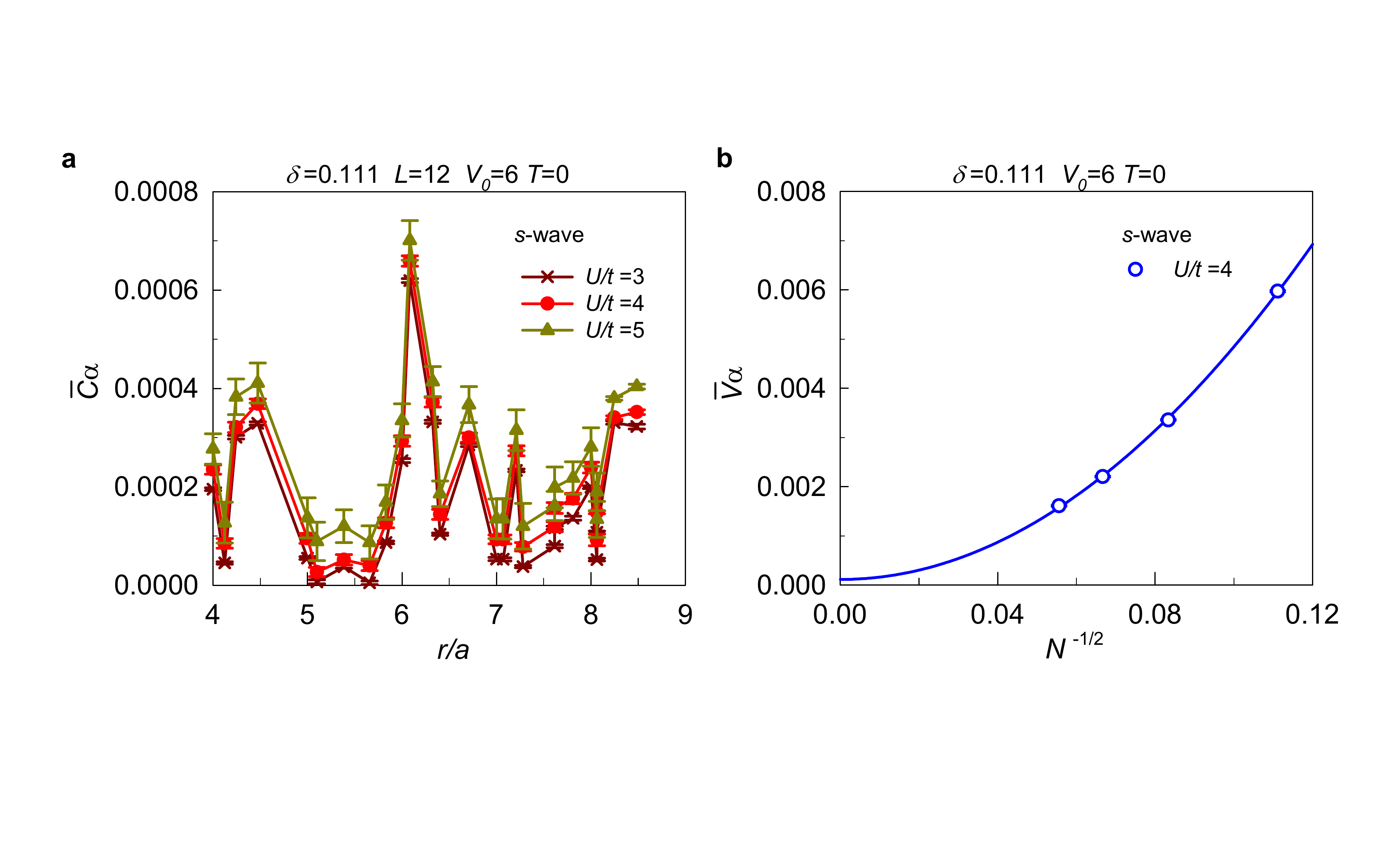}
\caption{(a) CPQMC-calculated vertex contributions of $s$ pairing symmetry $C_\alpha$ as
a function of distance $r$ with $\delta=0.111$, $T=0$, $V_0=6$ and $U/t=4$ at period $\mathcal{P}=3$ on a $L=12$ lattice
for the different $U/t$. (b) The scaling analysis is based on the average of the long-range $C_\alpha$. The solid blue line is fit on the third-order polynomial in $\frac{1}{\sqrt{N}}$.
}
\label{Fig:CPQMC}
\end{figure}

As shown in Fig.~\ref{Fig:CPQMC} (a), the vertex contributions of $s$-wave pairing symmetry $\bar{C}_s (\mathbf{r})$ are slightly enhanced by the coulomb interaction, while the leading pairing symmetry does not change, indicating the importance of electronic correlation. To study more about the dominant $s$-wave superconducting correlation in the thermodynamic limit, we also analyze the evolution of $\bar C_s$ with increasing lattice size. Furthermore, we calculate the averaged long-range $s$-wave vertex contributions,
$\bar{V}_s=\frac{1}{\sqrt{N'}} \sum_{r/a>4}\bar{C}_s (\mathbf{r})$, where $N'$ is the total number of electronic pairs with $r/a>4$, and $a$ represents the lattice spacing.
In Fig.~\ref{Fig:CPQMC} (b), $\bar{V}_s$ is plotted as a function of $\frac{1}{\sqrt{N}}$ with $\delta=0.111$, $V_0=6$ and $U/t=4$. We can clearly notice that $\bar{V}_s$ exhibits a small but finite positive
value in the thermodynamic limit, suggesting the possible occurrence of long-range $s$-wave superconducting order under the investigated parameter region.

\begin{figure}[tbp]
  \includegraphics[scale=0.65,trim = 0 0 0 0, clip]{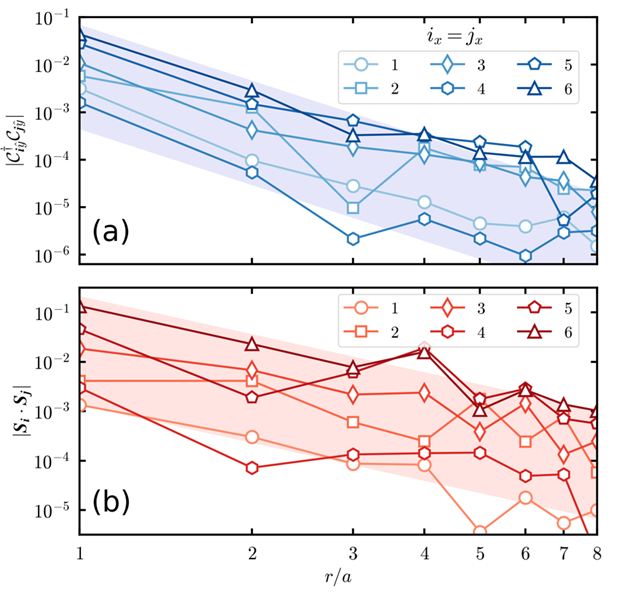}
  \caption{The DMRG-calculated (a) Cooper pair-pair correlation $\vert C_{\mathbf{i}y}^{\dagger} C_{\mathbf{j}y} \vert$ and (b) spin-spin correlation $\vert \mathbf{S}_\mathbf{i} \cdot \mathbf{S}_\mathbf{j} \vert$ as a function of distance $r=\vert \mathbf{i} - \mathbf{j} \vert$ in the $s$-wave state at $\delta=0.111$, $V_0 / t = U/t = 8$ and $\mathcal{P}=3$ on a $L_x \times L_y = 6 \times 18$ lattice. Here, we choose different values of $i_x = j_x = 1$, $2$, $3$, $4$, $5$, and $6$. The color-shaded regions indicate the linear tendency of the correlation functions in the double-logarithmic plot. $10240$ SU($2$) bases are used in DMRG, which is equivalent to $30000$ U($1$) bases roughly.
}
\label{Fig:dmrg}
\end{figure}

We also pay much afford to directly examine the decaying behavior of correlations with distance, with the aim of determining which, if any, dominates in the $s$-wave state.
In Figs.~\ref{Fig:dmrg}, we illustrate the $y$-axis of a $L_x \times L_y = 6 \times 18$ lattice as an example. The superconducting correlation $\vert C_{\mathbf{i}y}^{\dagger} C_{\mathbf{j}y} \vert$, which is also known as the Cooper pair-pair correlation, displays a quasi-long-range behavior that is dominated by a power-law when different choices of $i_x = j_x$ are made.
Additionally, we find that the superconducting correlations are identical for bonds situated along the circumference of the cylinder in comparison to those along the long side, aligning with the anticipated characteristics of the $s$-wave pairing. As shown in Fig.~\ref{Fig:dmrg}(b), the spin-spin correlations $\vert \mathbf{S}_\mathbf{i} \cdot \mathbf{S}_\mathbf{j} \vert$ exhibits quasi-long-ranged behavior, corresponding to a gapless spectrum in the spin sector, standing in sharp contrast to the $d$-wave stated in the recent study~\cite{science.aal5304}.
This finding indicates that the antiferromagnetic spin order and the superconducting order coexist.

\begin{center}
  \textbf{8. Spectrum of the two-particle density matrix}
\end{center}

\begin{figure}[tbp]
\includegraphics[scale=0.5,trim = 0 10 0 10, clip]{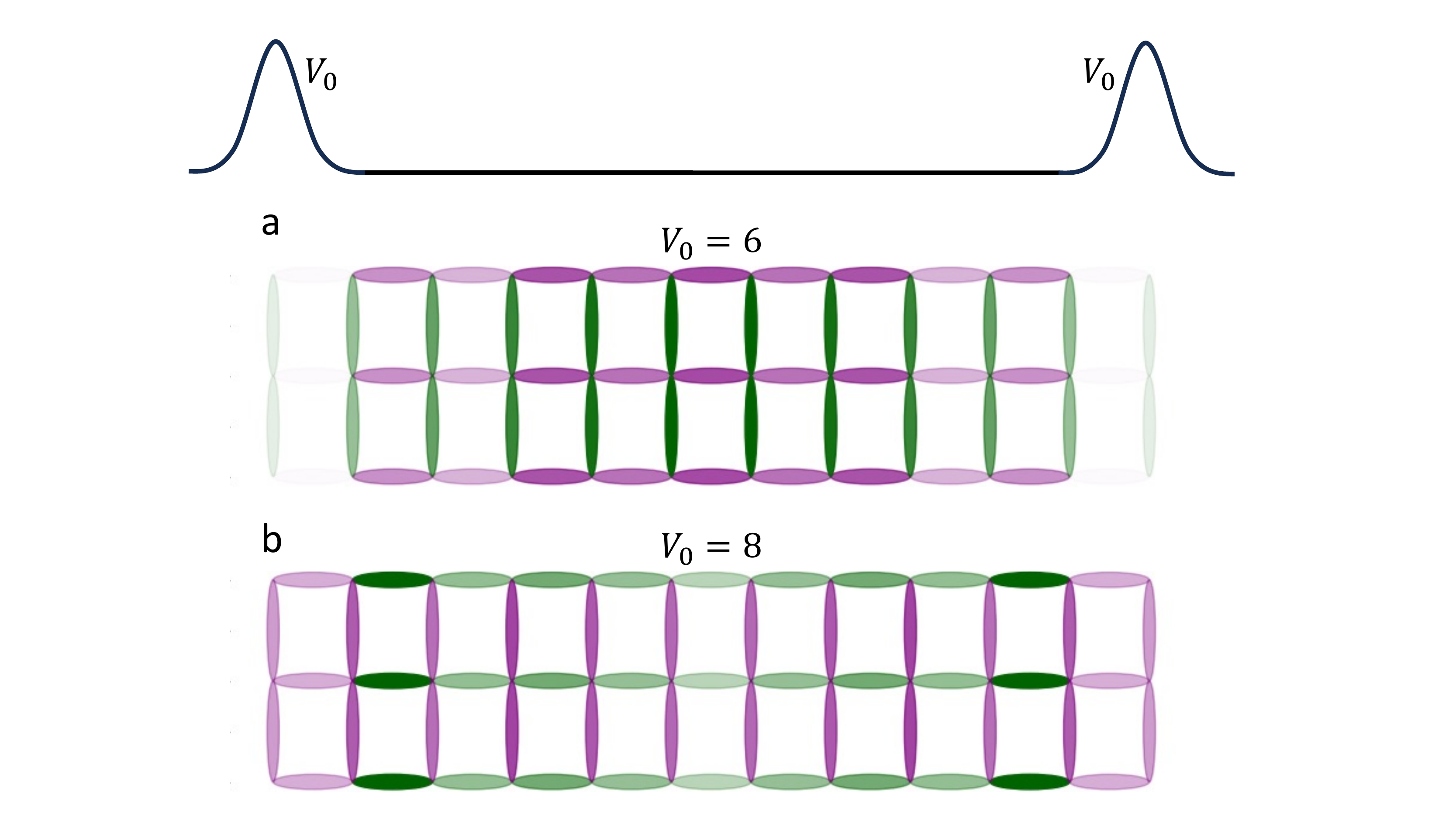}
\caption{Decomposition of two-particle density matrix \tck{(see Method)} for the $L_x \times L_y = 12 \times 3$ cylinder when the repulsion strength $U/t=8$ and \tck{hole-doping concentration} $\delta=1/9\approx0.111$ are both fixed. The \tck{charge-stripe amplitude} $V_0$ is performed on both ends of the lattice. The bond-distribution of \tck{condensate wave function} $\zeta_0 (\mathbf{i}\bm{\delta}_l)$ for the dominant \tck{Cooper pair mode} with \tck{(a) $V_0=6$ and (b) $V_0=8$}.
\tck{Purple (green) bonds indicate positive (negative) values of $\zeta_0 (\mathbf{i}\bm{\delta}_l)$.
It is noticed that $\zeta_0 (\mathbf{i}\bm{\delta}_l)$ has been normalized by the maximum value of $\vert \zeta_0 (\mathbf{i}\bm{\delta}_l) \vert$.}
The bond dimension in the SU($2$) DMRG \tck{$m=4096$} is large enough for the cylinder with the circumference width $L_y=3$.
}
\label{Fig:PairingCompetition}
\end{figure}

In Fig.~\ref{Fig:PairingCompetition},
we choose a long and thin cylinder with $L_x = 12 \gg L_y = 3$, and \tck{two charge stripes are situated} at \tck{the} edge circumferences, \tck{whose formation is beneficial for} extracting the location of distinct pairing symmetries in the real space.
For the charge-stripe amplitude $V_0=6$, in the dominant \tck{Cooper pair} mode $n=0$ with an eigenvalue of $\lambda_0 \approx 0.104$, the $d$-wave pairing is dominant \tck{in the inter-stripe region of the cylinder but weak near the on-stripe region}, which is characterized by a positive coefficient of \tck{$\zeta_0 (\mathbf{i} \bm{\delta}_l)$} on the horizontal bonds while negative one on the vertical bonds.
\tck{However, for large \tck{$V_0=8$}, the dominant pairing mode \tck{shows} extra strong extended $s$-wave pairing around the on-stripe region.}

Therefore, the extended $s$-wave pairing around the on-stripe region competes with the $d$-wave pairing inside the inter-stripe region far away from the domain wall, as shown in Fig.~4 of the main text.
In the two-dimensional thermodynamic limit of $L_x / L_y \approx 1$, the mentioned competition still exists and the analysis based on the robust DQMC calculations \tck{suggests that} a pairing-symmetry transition \tck{occurs} when $\mathcal{P} = 3$.

\begin{center}
  \textbf{9. Spin susceptibility}
\end{center}

\begin{figure}[h]
\includegraphics[scale=0.54,trim = 0 50 0 10, clip]{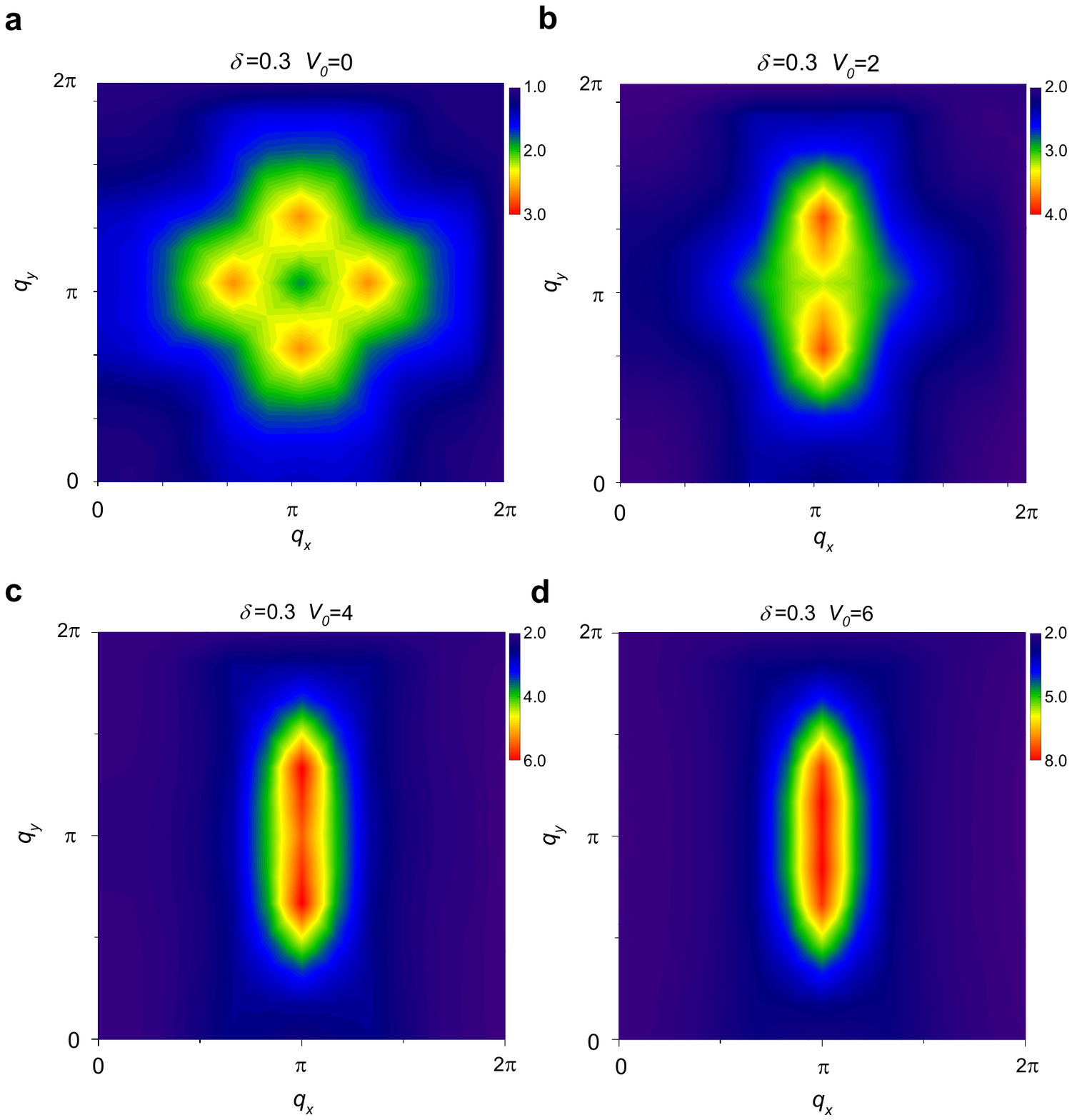}
\caption{(Color online) DQMC-calculated spin susceptibility $\chi_{s}(\mathbf{q})$ in the first Brillouin zone at $\delta=0.3$, $T=t/5$, \tck{$U/t=4$} for the different stripe potential of (a) $V_0$=0, (b) $V_0$=2, (c) $V_0$=4, (d) $V_0$=6.
}
\label{Fig:sus0.3}
\end{figure}

\begin{figure}[h]
\includegraphics[scale=0.54,trim = 0 50 0 30, clip]{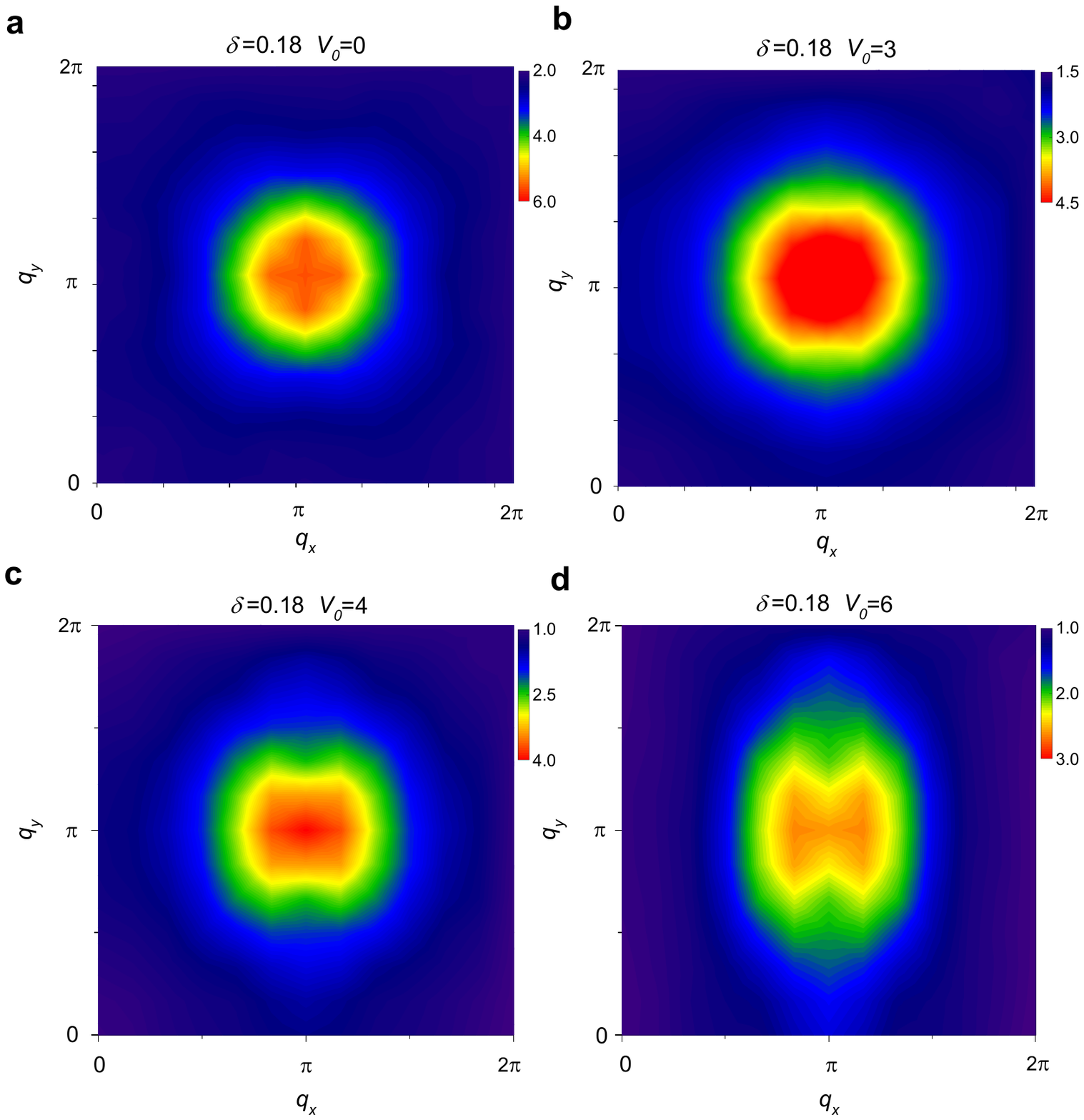}
\caption{(Color online) DQMC-calculated spin susceptibility $\chi_{s}(\mathbf{q})$ in the first Brillouin zone at $\delta=0.18$, $T=t/5$, $U/t=4$ for the different stripe potential of (a) $V_0$=0, (b) $V_0$=3, (c) $V_0$=4, (d) $V_0$=6.
}
\label{Fig:sus0.18}
\end{figure}

In Fig.~\ref{Fig:sus0.3}, we compare the obtained spin susceptibility $\chi_{s}(\mathbf{q})$ for $V_0=0 \sim 6$ at $\delta=0.3$, $T=t/5$, $U/t=4$ with striped period $\mathcal{P}=3$. We can notice that the $(\pi,\pi)$ magnetic correlation is enhanced as $V_0$ increases.
The system exhibits a strong antiferromagnetic correlation along the direction of stripes for $V_0=4\sim6$, which may be caused by the appearance of more nearly half-filled regions.
As for the vertical direction, magnetic correlation displays a rod-like feature in the first Brillouin zone.

\tck{Figure}~\ref{Fig:sus0.18} shows $\chi_{s}(\mathbf{q})$ in the first Brillouin zone at $\delta=0.18$, $T=t/5$, $U/t=4$ for the different $V_0$ in the $s$-wave region. Compared \tck{to} $\delta=0.3$, the $\chi_{s}(\mathbf{q})$ at $\delta=0.18$ displays different magnetic characteristics.
The magnetic correlation does not show a strong antiferromagnetic correlation along the direction of stripes as the $V_0$ increases. \tck{However,} the magnetic correlation has a significant change in the vertical direction and appears as a dumbbell pattern at large $V_0$.

\begin{center}
  \textbf{10. Temperature dependence of $\chi_{s}(\mathbf{q})$ and $\bar{P}_{\alpha}$}
\end{center}

\begin{figure}[tbp]
\includegraphics[scale=0.4,trim = 100 10 150 10, clip]{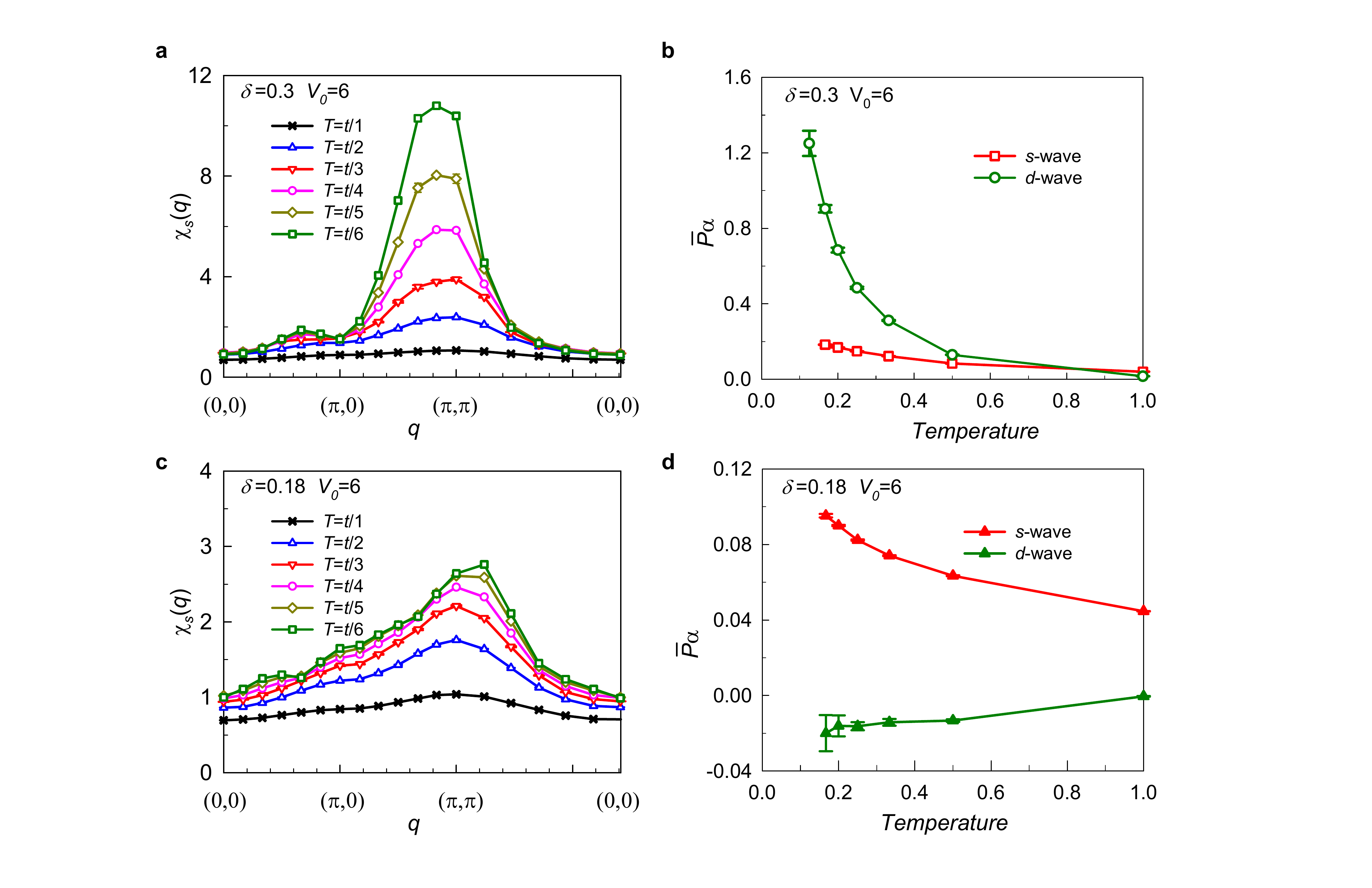}
\caption{(a) DQMC-calculated magnetic susceptibility $\chi_{s}(\mathbf{q})$ versus
momentum $q$ for different temperature at $\delta=0.3$, $V_0=6$, $U/t=4$ and $L=12$, (b) DQMC-calculated effective pairing interaction $\bar{P}_{\alpha}$ as a function of temperature at $\delta=0.3$, $V_0=6$, $U/t=4$ and $L=12$. (c)-(d) are similar to (a)-(b) but for the cases of $\delta=0.18$.
}
\label{Fig:suscep}
\end{figure}

We have studied side by side the temperature dependence
of \tck{the} spin susceptibility together with the pairing susceptibility in
Figs.~\ref{Fig:suscep}. And, we have also discussed the \tck{relationship} between spin fluctuation and superconductivity.
As shown in Figs.~\ref{Fig:suscep} (a) and (b) for the case of $\delta=0.3$, we
can \tck{see} that the $(\pi,\pi)$ magnetic correlation is quickly enhanced as the
temperature is lowered, i.e., the system exhibits a stronger AFM fluctuation.
Meanwhile, the dominant $d$ pairing symmetry is also enhanced with the decrease
of temperature. In Figs.~\ref{Fig:suscep} (c) and (d) for the case of $\delta=0.18$, both the $(\pi,\pi)$ magnetic correlation and robust $\bar{P}_{s}$ slowly increase as the temperature is lowered. From the results of $\delta=0.3$ and $\delta=0.18$, we can \tck{see} that the $(\pi,\pi)$ magnetic correlation and dominant pairing correlation exhibit a very similar temperature dependence. This strongly reveals that spin fluctuation and superconductivity are strongly interwoven.

\begin{center}
  \textbf{11. Cosine-like charge modulation}
\end{center}

\begin{figure}[tbp]
\includegraphics[scale=0.4,trim = 140 50 50 25, clip]{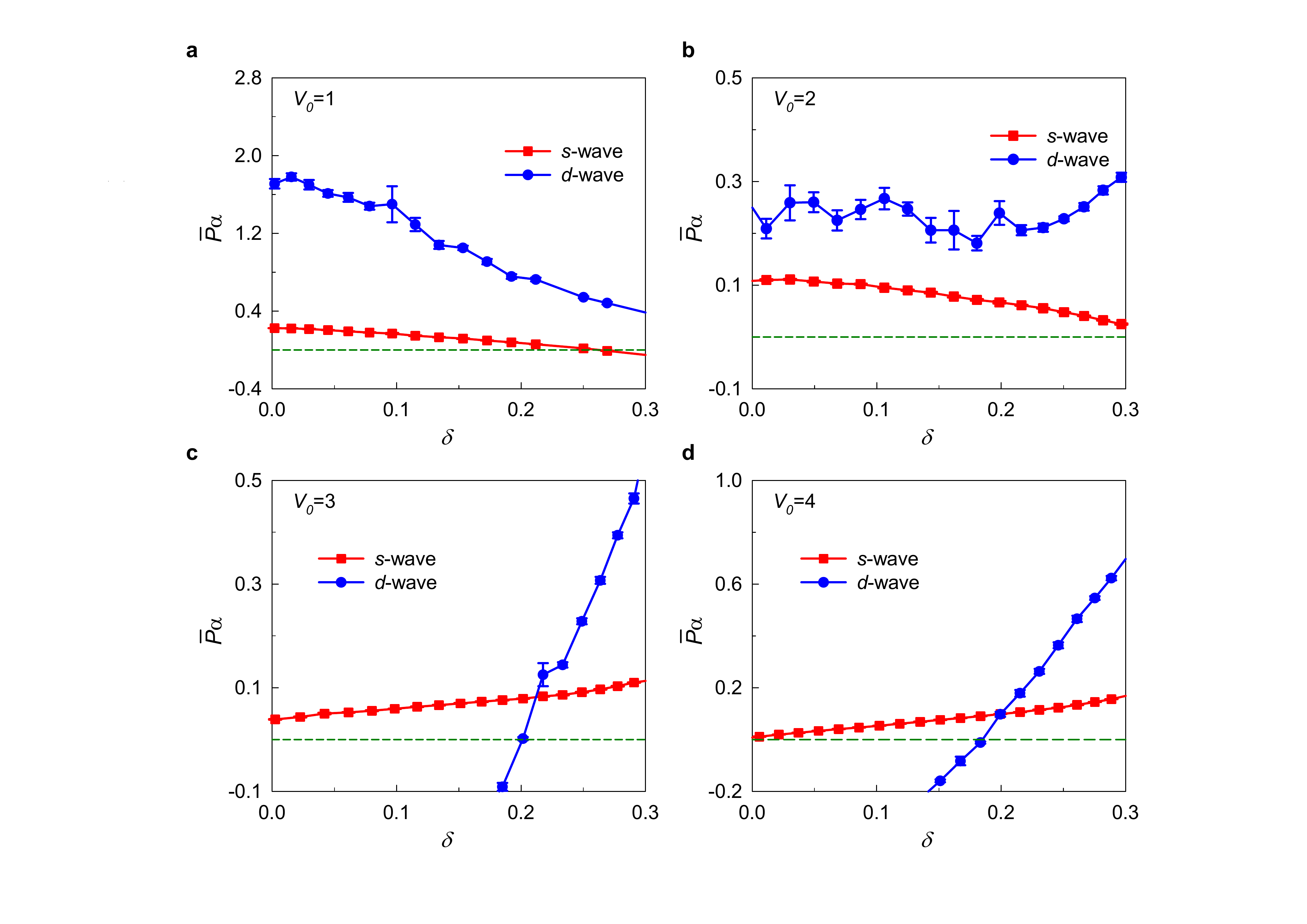}
\caption{DQMC-calculated effective pairing interaction $\bar{P}_{\alpha}$ as
a function of hole-doping concentration $\delta$ at $T=t/5$ and $U/t=4$ with
cosine-like stripe potential at period $\mathcal{P}=3$ on a $L=12$ lattice
for the different stripe potential (a) $V_0$=1, (b) $V_0$=2, (c) $V_0$=3,
and (d) $V_0$=4.
}
\label{Fig:cos}
\end{figure}

We also choose a cosine-like varying charge modulation
$V(l_x)$ with magnitude $V_0$ for period $\mathcal{P}=3$.
Specifically, $V(l_x)$ appears as \tck{$V_0\cos(0)$, $V_0\cos(2\pi/3)$,
$V_0\cos(4\pi/3)$ and $V_0\cos(6\pi/3)$}. Thus, the period of cosine-like
potential is $\mathcal{P}=3$.
For $V_0=1\sim2$ in \tck{Figs.~\ref{Fig:cos}(a) and (b)}, we can notice that
the $\bar{P}_{d}$ is always dominant under all hole-doping \tck{concentrations}.
Interestingly, in \tck{Fig.~\ref{Fig:cos}(c)}, when \tck{$V_0=3$}, the $\bar{P}_{s}$
eventually becomes more stable than $\bar{P}_{d}$ under a large $\delta$
range ($0<\delta \leq0.2$), leading to a novel $d$-$s$ wave transition.
Our major conclusion is independent of the different styles of added stripes but mainly depends on the period of \tck{the stripes}. In addition, from our simulations, we can easily get that the critical value of
$V_0$ for \tck{the} $d$-$s$ wave transition becomes smaller when we choose a
cosine-like varying modulation.

\begin{center}
  \textbf{12. Sign problem of DQMC}
\end{center}

\begin{figure}[h]
\includegraphics[scale=0.44,trim = 50 140 0 250, clip]{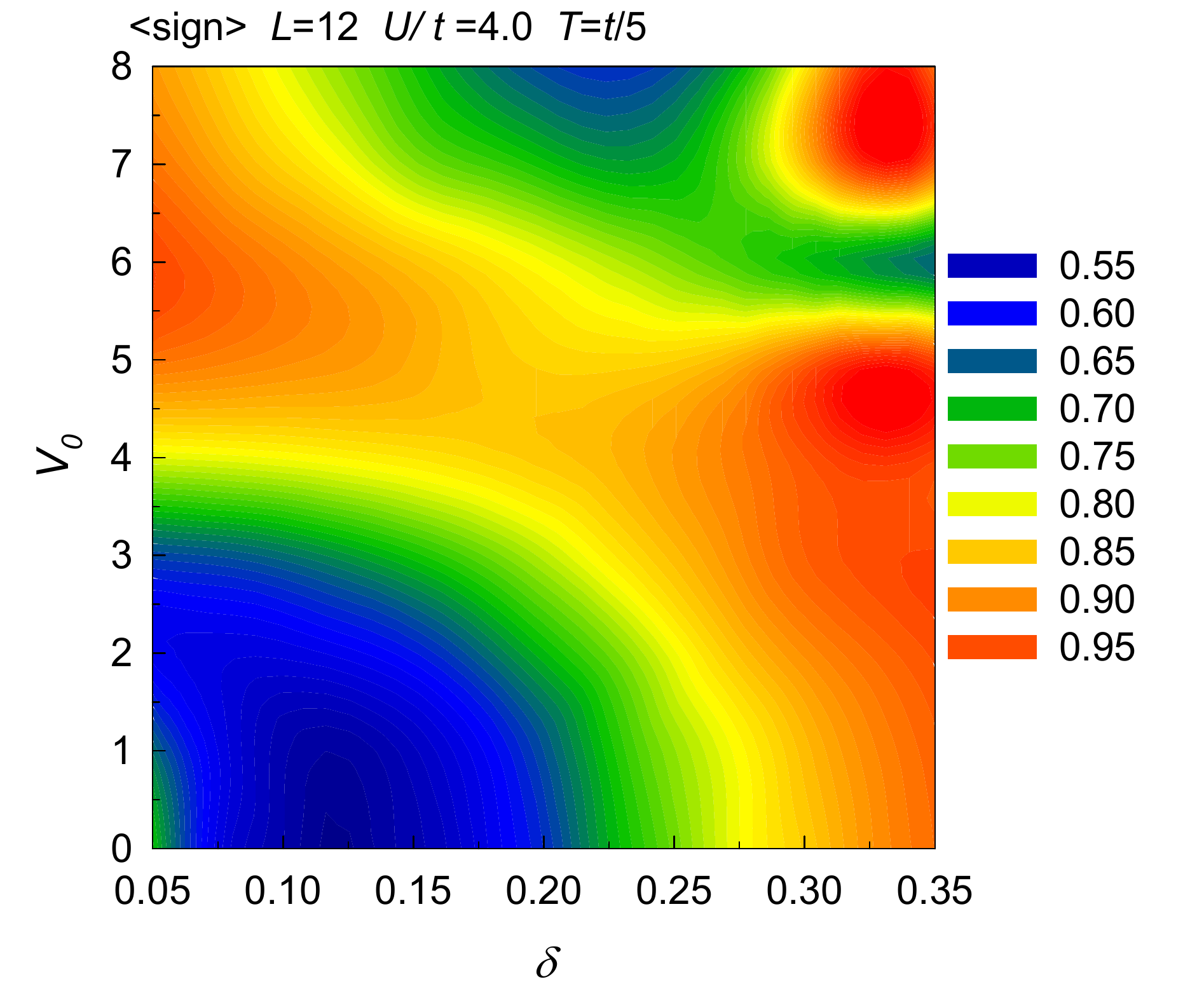}
\caption{(Color online) Contour map of the average sign $\avg{sign}$ as a function of $V_0$ and $\delta$ at $U/t=4$ and $T=t/5$ on a $L=12$ square lattice.
}
\label{Fig:sign2}
\end{figure}

In Fig.~\ref{Fig:sign2}, we show a contour map of $\avg{\text{sign}}$ as a function of $V_0$ and $\delta$ at \tck{$U/t=4$} and $T=t/5$.
Although the \tck{charge-stripe amplitude} $V_0$ breaks \tck{the} particle-hole symmetry, we can see that the sign problem does not get worse in the doped region with $V_0=4\sim7$. Particularly, our simulations are mainly performed at doping levels \tck{of} $\delta=0.3$ and \tck{$0.18$}, where the sign problem is relatively tolerated.

\begin{figure}[tbp]
\includegraphics[scale=0.36,trim = 120 370 150 90, clip]{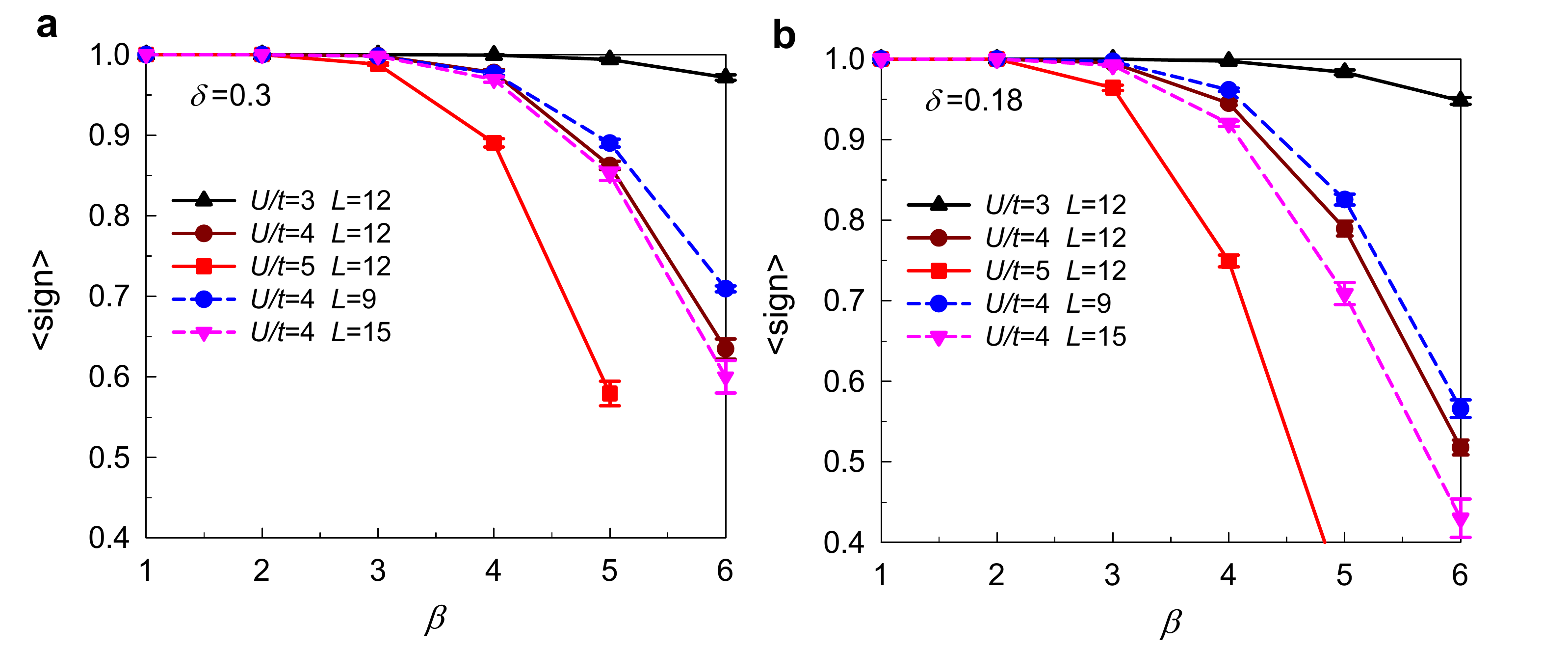}
\caption{(Color online) Average sign $\avg{sign}$ versus the inverse temperatures $\beta=1/T$
for different $U/t$ and lattice sizes at (a) $\delta=0.3$, (b) $\delta=0.18$.
}
\label{Fig:sign}
\end{figure}

Besides, we also perform an analysis \tck{of} the infamous sign problem within the parameter range investigated. In Fig.~\ref{Fig:sign},
the average sign decreases quickly as the inverse temperature exceeds $3$.
We can \tck{see} that the sign problem becomes worse for higher interaction or larger lattice \tck{sizes}. To guarantee \tck{the} same quality of data with $\avg{sign}\approx1$, much longer measurements are performed to compensate the fluctuations when \tck{the} sign problem is severe. Actually, the runs should be amplified by a factor on the order of $\avg{sign}^{-2}$\cite{PhysRevD.24.2278,SANTOS2003}. Therefore, some of the simulations are performed with more than 1,200,000 measurements.
These efforts \tck{make} our results reliable.

\newpage
\bibliography{reference}

\begin{thebibliography}{10}
\expandafter\ifx\csname url\endcsname\relax
  \def\url#1{\texttt{#1}}\fi
\expandafter\ifx\csname urlprefix\endcsname\relax\def\urlprefix{URL }\fi
\providecommand{\bibinfo}[2]{#2}
\providecommand{\eprint}[2][]{\url{#2}}

\bibitem{Tranquada1995}
\bibinfo{author}{Tranquada, J.~M.}, \bibinfo{author}{Sternlieb, B.~J.},
  \bibinfo{author}{Axe, J.~D.}, \bibinfo{author}{Nakamura, Y.} \&
  \bibinfo{author}{Uchida, S.}
\newblock \bibinfo{title}{Evidence for stripe correlations of spins and holes
  in copper oxide superconductors}.
\newblock \emph{\bibinfo{journal}{Nature}} \textbf{\bibinfo{volume}{375}},
  \bibinfo{pages}{561--563} (\bibinfo{year}{1995}).

\bibitem{Abbamonte2005}
\bibinfo{author}{Abbamonte, P.} \emph{et~al.}
\newblock \bibinfo{title}{Spatially modulated ``mottness'' in
  $\mathrm{La}_{2-x}\mathrm{Ba}_{x}\mathrm{Cu}\mathrm{O}_{4}$}.
\newblock \emph{\bibinfo{journal}{Nature Physics}}
  \textbf{\bibinfo{volume}{1}}, \bibinfo{pages}{155--158}
  (\bibinfo{year}{2005}).

\bibitem{science.1223532}
\bibinfo{author}{Ghiringhelli, G.} \emph{et~al.}
\newblock \bibinfo{title}{Long-range incommensurate charge fluctuations in
  ($\mathrm{Y}$,$\mathrm{Nd}$)$\mathrm{Ba}_2$$\mathrm{Cu}_3$$\mathrm{O}_{6+x}$}.
\newblock \emph{\bibinfo{journal}{Science}} \textbf{\bibinfo{volume}{337}},
  \bibinfo{pages}{821--825} (\bibinfo{year}{2012}).

\bibitem{science.1243479}
\bibinfo{author}{da~Silva~Neto, E.~H.} \emph{et~al.}
\newblock \bibinfo{title}{Ubiquitous interplay between charge ordering and
  high-temperature superconductivity in cuprates}.
\newblock \emph{\bibinfo{journal}{Science}} \textbf{\bibinfo{volume}{343}},
  \bibinfo{pages}{393--396} (\bibinfo{year}{2014}).

\bibitem{Fradkin2012}
\bibinfo{author}{Fradkin, E.} \& \bibinfo{author}{Kivelson, S.~A.}
\newblock \bibinfo{title}{Ineluctable complexity}.
\newblock \emph{\bibinfo{journal}{Nature Physics}}
  \textbf{\bibinfo{volume}{8}}, \bibinfo{pages}{864--866}
  (\bibinfo{year}{2012}).

\bibitem{Wang2016}
\bibinfo{author}{Wang, Q.} \emph{et~al.}
\newblock \bibinfo{title}{Strong interplay between stripe spin fluctuations,
  nematicity and superconductivity in fese}.
\newblock \emph{\bibinfo{journal}{Nature Materials}}
  \textbf{\bibinfo{volume}{15}}, \bibinfo{pages}{159--163}
  (\bibinfo{year}{2016}).

\bibitem{Gu2020}
\bibinfo{author}{Gu, Q.} \emph{et~al.}
\newblock \bibinfo{title}{Single particle tunneling spectrum of superconducting
  $\mathrm{Nd_{1-x}Sr_{x}NiO{_2}}$ thin films}.
\newblock \emph{\bibinfo{journal}{Nature Communications}}
  \textbf{\bibinfo{volume}{11}}, \bibinfo{pages}{6027} (\bibinfo{year}{2020}).

\bibitem{Wang2021}
\bibinfo{author}{Wang, B.~Y.} \emph{et~al.}
\newblock \bibinfo{title}{Isotropic pauli-limited superconductivity in the
  infinite-layer nickelate $\mathrm{Nd{_{0.775}}Sr{_{0.225}}NiO{_2}}$}.
\newblock \emph{\bibinfo{journal}{Nature Physics}}
  \textbf{\bibinfo{volume}{17}}, \bibinfo{pages}{473--477}
  (\bibinfo{year}{2021}).

\bibitem{arxiv.2201.12971}
\bibinfo{author}{Harvey, S.~P.} \emph{et~al.}
\newblock \bibinfo{title}{Evidence for nodal superconductivity in
  infinite-layer nickelates} (\bibinfo{year}{2022}).
\newblock \eprint{arXiv: 2201.12971}.

\bibitem{arXiv:2201.10038}
\bibinfo{author}{Chow, L.~E.} \emph{et~al.}
\newblock \bibinfo{title}{Pairing symmetry in infinite-layer nickelate
  superconductor} (\bibinfo{year}{2022}).
\newblock \eprint{arXiv: 2201.10038}.

\bibitem{Ji2023}
\bibinfo{author}{Ji, H.} \emph{et~al.}
\newblock \bibinfo{title}{Rotational symmetry breaking in superconducting
  nickelate $\mathrm{Nd_{0.8}Sr_{0.2}NiO{_2}}$ films}.
\newblock \emph{\bibinfo{journal}{Nature Communications}}
  \textbf{\bibinfo{volume}{14}}, \bibinfo{pages}{7155} (\bibinfo{year}{2023}).

\bibitem{Cheng2024}
\bibinfo{author}{Cheng, B.} \emph{et~al.}
\newblock \bibinfo{title}{Evidence for d-wave superconductivity of
  infinite-layer nickelates from low-energy electrodynamics}.
\newblock \emph{\bibinfo{journal}{Nature Materials}}  (\bibinfo{year}{2024}).

\bibitem{PhysRevB.61.R854}
\bibinfo{author}{Yoshizawa, H.} \emph{et~al.}
\newblock \bibinfo{title}{Stripe order at low temperatures in
  $\mathrm{La}_{2-x}\mathrm{Sr}_{x}\mathrm{NiO}_{4}$ with
  $0.289\ensuremath{\lesssim}x\ensuremath{\lesssim}0.5$}.
\newblock \emph{\bibinfo{journal}{Phys. Rev. B}} \textbf{\bibinfo{volume}{61}},
  \bibinfo{pages}{R854--R857} (\bibinfo{year}{2000}).

\bibitem{Zheng2022}
\bibinfo{author}{Zheng, L.} \emph{et~al.}
\newblock \bibinfo{title}{Emergent charge order in pressurized kagome
  superconductor $\mathrm{CsV_{3}Sb{_5}}$}.
\newblock \emph{\bibinfo{journal}{Nature}} \textbf{\bibinfo{volume}{611}},
  \bibinfo{pages}{682--687} (\bibinfo{year}{2022}).

\bibitem{Liu2024}
\bibinfo{author}{Liu, Y.} \emph{et~al.}
\newblock \bibinfo{title}{Superconductivity under pressure in a chromium-based
  kagome metal}.
\newblock \emph{\bibinfo{journal}{Nature}} \textbf{\bibinfo{volume}{632}},
  \bibinfo{pages}{1032--1037} (\bibinfo{year}{2024}).

\bibitem{Ding2023}
\bibinfo{author}{Ding, X.} \emph{et~al.}
\newblock \bibinfo{title}{Critical role of hydrogen for superconductivity in
  nickelates}.
\newblock \emph{\bibinfo{journal}{Nature}} \textbf{\bibinfo{volume}{615}},
  \bibinfo{pages}{50--55} (\bibinfo{year}{2023}).

\bibitem{arXiv:2306.15086}
\bibinfo{author}{Pelliciari, J.} \emph{et~al.}
\newblock \bibinfo{title}{Comment on newly found charge density waves in
  infinite layer nickelates} (\bibinfo{year}{2023}).
\newblock \eprint{arXiv: 2306.15086}.

\bibitem{arXiv:2307.13569}
\bibinfo{author}{Tam, C.~C.} \emph{et~al.}
\newblock \bibinfo{title}{Reply to ``comment on newly found charge density
  waves in infinite layer nickelates''} (\bibinfo{year}{2023}).
\newblock \eprint{arXiv: 2307.13569}.

\bibitem{Parzyck2024}
\bibinfo{author}{Parzyck, C.~T.} \emph{et~al.}
\newblock \bibinfo{title}{Absence of $3a_0$ charge density wave order in the
  infinite-layer nickelate $\mathrm{Nd} \mathrm{NiO}_{2}$}.
\newblock \emph{\bibinfo{journal}{Nature Materials}}
  \textbf{\bibinfo{volume}{23}}, \bibinfo{pages}{486--491}
  (\bibinfo{year}{2024}).

\bibitem{Keimer2015}
\bibinfo{author}{Keimer, B.}, \bibinfo{author}{Kivelson, S.~A.},
  \bibinfo{author}{Norman, M.~R.}, \bibinfo{author}{Uchida, S.} \&
  \bibinfo{author}{Zaanen, J.}
\newblock \bibinfo{title}{From quantum matter to high-temperature
  superconductivity in copper oxides}.
\newblock \emph{\bibinfo{journal}{Nature}} \textbf{\bibinfo{volume}{518}},
  \bibinfo{pages}{179--186} (\bibinfo{year}{2015}).

\bibitem{annurev102024}
\bibinfo{author}{Arovas, D.~P.}, \bibinfo{author}{Berg, E.},
  \bibinfo{author}{Kivelson, S.~A.} \& \bibinfo{author}{Raghu, S.}
\newblock \bibinfo{title}{The hubbard model}.
\newblock \emph{\bibinfo{journal}{Annual Review of Condensed Matter Physics}}
  \textbf{\bibinfo{volume}{13}}, \bibinfo{pages}{239--274}
  (\bibinfo{year}{2022}).

\bibitem{RevModPhys.66.763}
\bibinfo{author}{Dagotto, E.}
\newblock \bibinfo{title}{Correlated electrons in high-temperature
  superconductors}.
\newblock \emph{\bibinfo{journal}{Rev. Mod. Phys.}}
  \textbf{\bibinfo{volume}{66}}, \bibinfo{pages}{763--840}
  (\bibinfo{year}{1994}).

\bibitem{RevModPhys.84.1383}
\bibinfo{author}{Scalapino, D.~J.}
\newblock \bibinfo{title}{A common thread: The pairing interaction for
  unconventional superconductors}.
\newblock \emph{\bibinfo{journal}{Rev. Mod. Phys.}}
  \textbf{\bibinfo{volume}{84}}, \bibinfo{pages}{1383--1417}
  (\bibinfo{year}{2012}).

\bibitem{RevModPhys.87.457}
\bibinfo{author}{Fradkin, E.}, \bibinfo{author}{Kivelson, S.~A.} \&
  \bibinfo{author}{Tranquada, J.~M.}
\newblock \bibinfo{title}{Colloquium: Theory of intertwined orders in high
  temperature superconductors}.
\newblock \emph{\bibinfo{journal}{Rev. Mod. Phys.}}
  \textbf{\bibinfo{volume}{87}}, \bibinfo{pages}{457--482}
  (\bibinfo{year}{2015}).

\bibitem{science.aam7127}
\bibinfo{author}{Zheng, B.-X.} \emph{et~al.}
\newblock \bibinfo{title}{Stripe order in the underdoped region of the
  two-dimensional hubbard model}.
\newblock \emph{\bibinfo{journal}{Science}} \textbf{\bibinfo{volume}{358}},
  \bibinfo{pages}{1155--1160} (\bibinfo{year}{2017}).

\bibitem{science.aak9546}
\bibinfo{author}{Huang, E.~W.} \emph{et~al.}
\newblock \bibinfo{title}{Numerical evidence of fluctuating stripes in the
  normal state of high-$\mathrm{Tc}$ cuprate superconductors}.
\newblock \emph{\bibinfo{journal}{Science}} \textbf{\bibinfo{volume}{358}},
  \bibinfo{pages}{1161--1164} (\bibinfo{year}{2017}).

\bibitem{PhysRevB.35.3359}
\bibinfo{author}{Lin, H.~Q.} \& \bibinfo{author}{Hirsch, J.~E.}
\newblock \bibinfo{title}{Two-dimensional hubbard model with nearest- and
  next-nearest-neighbor hopping}.
\newblock \emph{\bibinfo{journal}{Phys. Rev. B}} \textbf{\bibinfo{volume}{35}},
  \bibinfo{pages}{3359--3368} (\bibinfo{year}{1987}).

\bibitem{PhysRevB.62.R9283}
\bibinfo{author}{Lichtenstein, A.~I.} \& \bibinfo{author}{Katsnelson, M.~I.}
\newblock \bibinfo{title}{Antiferromagnetism and $d$-wave superconductivity in
  cuprates: A cluster dynamical mean-field theory}.
\newblock \emph{\bibinfo{journal}{Phys. Rev. B}} \textbf{\bibinfo{volume}{62}},
  \bibinfo{pages}{R9283--R9286} (\bibinfo{year}{2000}).

\bibitem{npjQM.3.22}
\bibinfo{author}{Huang, E.~W.}, \bibinfo{author}{Mendl, C.~B.},
  \bibinfo{author}{Jiang, H.-C.}, \bibinfo{author}{Moritz, B.} \&
  \bibinfo{author}{Devereaux, T.~P.}
\newblock \bibinfo{title}{Stripe order from the perspective of the hubbard
  model}.
\newblock \emph{\bibinfo{journal}{npj Quantum Materials}}
  \textbf{\bibinfo{volume}{3}}, \bibinfo{pages}{22} (\bibinfo{year}{2018}).

\bibitem{PhysRevLett.94.156404}
\bibinfo{author}{S\'en\'echal, D.}, \bibinfo{author}{Lavertu, P.-L.},
  \bibinfo{author}{Marois, M.-A.} \& \bibinfo{author}{Tremblay, A.-M.~S.}
\newblock \bibinfo{title}{Competition between antiferromagnetism and
  superconductivity in high-${T}_{c}$ cuprates}.
\newblock \emph{\bibinfo{journal}{Phys. Rev. Lett.}}
  \textbf{\bibinfo{volume}{94}}, \bibinfo{pages}{156404}
  (\bibinfo{year}{2005}).

\bibitem{PhysRevLett.104.247001}
\bibinfo{author}{Maier, T.~A.}, \bibinfo{author}{Alvarez, G.},
  \bibinfo{author}{Summers, M.} \& \bibinfo{author}{Schulthess, T.~C.}
\newblock \bibinfo{title}{Dynamic cluster quantum monte carlo simulations of a
  two-dimensional hubbard model with stripelike charge-density-wave
  modulations: Interplay between inhomogeneities and the superconducting
  state}.
\newblock \emph{\bibinfo{journal}{Phys. Rev. Lett.}}
  \textbf{\bibinfo{volume}{104}}, \bibinfo{pages}{247001}
  (\bibinfo{year}{2010}).

\bibitem{science.adh7691}
\bibinfo{author}{Xu, H.} \emph{et~al.}
\newblock \bibinfo{title}{Coexistence of superconductivity with partially
  filled stripes in the hubbard model}.
\newblock \emph{\bibinfo{journal}{Science}} \textbf{\bibinfo{volume}{384}},
  \bibinfo{pages}{eadh7691} (\bibinfo{year}{2024}).

\bibitem{PhysRevB.86.184506}
\bibinfo{author}{Mondaini, R.}, \bibinfo{author}{Ying, T.},
  \bibinfo{author}{Paiva, T.} \& \bibinfo{author}{Scalettar, R.~T.}
\newblock \bibinfo{title}{Determinant quantum monte carlo study of the
  enhancement of $d$-wave pairing by charge inhomogeneity}.
\newblock \emph{\bibinfo{journal}{Phys. Rev. B}} \textbf{\bibinfo{volume}{86}},
  \bibinfo{pages}{184506} (\bibinfo{year}{2012}).

\bibitem{pnas.2109406119}
\bibinfo{author}{Jiang, H.-C.} \& \bibinfo{author}{Kivelson, S.~A.}
\newblock \bibinfo{title}{Stripe order enhanced superconductivity in the
  hubbard model}.
\newblock \emph{\bibinfo{journal}{Proceedings of the National Academy of
  Sciences}} \textbf{\bibinfo{volume}{119}}, \bibinfo{pages}{e2109406119}
  (\bibinfo{year}{2022}).

\bibitem{PhysRevB.72.060502}
\bibinfo{author}{Martin, I.}, \bibinfo{author}{Podolsky, D.} \&
  \bibinfo{author}{Kivelson, S.~A.}
\newblock \bibinfo{title}{Enhancement of superconductivity by local
  inhomogeneities}.
\newblock \emph{\bibinfo{journal}{Phys. Rev. B}} \textbf{\bibinfo{volume}{72}},
  \bibinfo{pages}{060502} (\bibinfo{year}{2005}).

\bibitem{10.1093nwae194}
\bibinfo{author}{Ding, X.} \emph{et~al.}
\newblock \bibinfo{title}{{Cuprate-like electronic structures in infinite-layer
  nickelates with substantial hole dopings}}.
\newblock \emph{\bibinfo{journal}{National Science Review}}
  \textbf{\bibinfo{volume}{11}}, \bibinfo{pages}{nwae194}
  (\bibinfo{year}{2024}).

\bibitem{arxiv.2403.07344}
\bibinfo{author}{Sun, W.} \emph{et~al.}
\newblock \bibinfo{title}{Electronic structure of superconducting
  infinite-layer lanthanum nickelates} (\bibinfo{year}{2024}).
\newblock \eprint{arXiv: 2403.07344}.

\bibitem{PhysRevLett.125.147003}
\bibinfo{author}{Zeng, S.} \emph{et~al.}
\newblock \bibinfo{title}{Phase diagram and superconducting dome of
  infinite-layer
  $\mathrm{Nd}_{1\ensuremath{-}x}\mathrm{Sr}_{x}\mathrm{NiO}_{2}$ thin films}.
\newblock \emph{\bibinfo{journal}{Phys. Rev. Lett.}}
  \textbf{\bibinfo{volume}{125}}, \bibinfo{pages}{147003}
  (\bibinfo{year}{2020}).

\bibitem{PhysRevLett.125.027001}
\bibinfo{author}{Li, D.} \emph{et~al.}
\newblock \bibinfo{title}{Superconducting dome in
  $\mathrm{Nd}_{1\ensuremath{-}x}\mathrm{Sr}_{x}\mathrm{NiO}_{2}$ infinite
  layer films}.
\newblock \emph{\bibinfo{journal}{Phys. Rev. Lett.}}
  \textbf{\bibinfo{volume}{125}}, \bibinfo{pages}{027001}
  (\bibinfo{year}{2020}).

\bibitem{Wu2011}
\bibinfo{author}{Wu, T.} \emph{et~al.}
\newblock \bibinfo{title}{Magnetic-field-induced charge-stripe order in the
  high-temperature superconductor $\mathrm{YBa{_2}Cu{_3}O{_y}}$}.
\newblock \emph{\bibinfo{journal}{Nature}} \textbf{\bibinfo{volume}{477}},
  \bibinfo{pages}{191--194} (\bibinfo{year}{2011}).

\bibitem{PhysRevX.6.021004}
\bibinfo{author}{Badoux, S.} \emph{et~al.}
\newblock \bibinfo{title}{Critical doping for the onset of fermi-surface
  reconstruction by charge-density-wave order in the cuprate superconductor
  $\mathrm{La}_{2\ensuremath{-}x}\mathrm{Sr}_{x}\mathrm{CuO}_{4}$}.
\newblock \emph{\bibinfo{journal}{Phys. Rev. X}} \textbf{\bibinfo{volume}{6}},
  \bibinfo{pages}{021004} (\bibinfo{year}{2016}).

\bibitem{PhysRevD.24.2278}
\bibinfo{author}{Blankenbecler, R.}, \bibinfo{author}{Scalapino, D.~J.} \&
  \bibinfo{author}{Sugar, R.~L.}
\newblock \bibinfo{title}{Monte carlo calculations of coupled boson-fermion
  systems. i}.
\newblock \emph{\bibinfo{journal}{Phys. Rev. D}} \textbf{\bibinfo{volume}{24}},
  \bibinfo{pages}{2278--2286} (\bibinfo{year}{1981}).

\bibitem{PhysRevB.40.506}
\bibinfo{author}{White, S.~R.} \emph{et~al.}
\newblock \bibinfo{title}{Numerical study of the two-dimensional hubbard
  model}.
\newblock \emph{\bibinfo{journal}{Phys. Rev. B}} \textbf{\bibinfo{volume}{40}},
  \bibinfo{pages}{506--516} (\bibinfo{year}{1989}).

\bibitem{PhysRevLett.120.116601}
\bibinfo{author}{Ma, T.}, \bibinfo{author}{Zhang, L.}, \bibinfo{author}{Chang,
  C.-C.}, \bibinfo{author}{Hung, H.-H.} \& \bibinfo{author}{Scalettar, R.~T.}
\newblock \bibinfo{title}{Localization of interacting dirac fermions}.
\newblock \emph{\bibinfo{journal}{Phys. Rev. Lett.}}
  \textbf{\bibinfo{volume}{120}}, \bibinfo{pages}{116601}
  (\bibinfo{year}{2018}).

\bibitem{PhysRevB.99.195147}
\bibinfo{author}{Zhang, L.}, \bibinfo{author}{Ma, T.}, \bibinfo{author}{Costa,
  N.~C.}, \bibinfo{author}{dos Santos, R.~R.} \& \bibinfo{author}{Scalettar,
  R.~T.}
\newblock \bibinfo{title}{Determinant quantum monte carlo study of exhaustion
  in the periodic anderson model}.
\newblock \emph{\bibinfo{journal}{Phys. Rev. B}} \textbf{\bibinfo{volume}{99}},
  \bibinfo{pages}{195147} (\bibinfo{year}{2019}).

\bibitem{science.abg9299}
\bibinfo{author}{Mondaini, R.}, \bibinfo{author}{Tarat, S.} \&
  \bibinfo{author}{Scalettar, R.~T.}
\newblock \bibinfo{title}{Quantum critical points and the sign problem}.
\newblock \emph{\bibinfo{journal}{Science}} \textbf{\bibinfo{volume}{375}},
  \bibinfo{pages}{418--424} (\bibinfo{year}{2022}).

\bibitem{PhysRevB.39.839}
\bibinfo{author}{White, S.~R.}, \bibinfo{author}{Scalapino, D.~J.},
  \bibinfo{author}{Sugar, R.~L.}, \bibinfo{author}{Bickers, N.~E.} \&
  \bibinfo{author}{Scalettar, R.~T.}
\newblock \bibinfo{title}{Attractive and repulsive pairing interaction vertices
  for the two-dimensional hubbard model}.
\newblock \emph{\bibinfo{journal}{Phys. Rev. B}} \textbf{\bibinfo{volume}{39}},
  \bibinfo{pages}{839--842} (\bibinfo{year}{1989}).

\bibitem{PhysRevLett.110.107002}
\bibinfo{author}{Ma, T.}, \bibinfo{author}{Lin, H.-Q.} \& \bibinfo{author}{Hu,
  J.}
\newblock \bibinfo{title}{Quantum monte carlo study of a dominant $s$-wave
  pairing symmetry in iron-based superconductors}.
\newblock \emph{\bibinfo{journal}{Phys. Rev. Lett.}}
  \textbf{\bibinfo{volume}{110}}, \bibinfo{pages}{107002}
  (\bibinfo{year}{2013}).

\bibitem{HUANG2019310}
\bibinfo{author}{Huang, T.}, \bibinfo{author}{Zhang, L.} \&
  \bibinfo{author}{Ma, T.}
\newblock \bibinfo{title}{Antiferromagnetically ordered mott insulator and
  $d$+i$d$ superconductivity in twisted bilayer graphene: a quantum monte carlo
  study}.
\newblock \emph{\bibinfo{journal}{Science Bulletin}}
  \textbf{\bibinfo{volume}{64}}, \bibinfo{pages}{310--314}
  (\bibinfo{year}{2019}).

\bibitem{annurev050711}
\bibinfo{author}{Agterberg, D.~F.} \emph{et~al.}
\newblock \bibinfo{title}{The physics of pair-density waves: Cuprate
  superconductors and beyond}.
\newblock \emph{\bibinfo{journal}{Annual Review of Condensed Matter Physics}}
  \textbf{\bibinfo{volume}{11}}, \bibinfo{pages}{231--270}
  (\bibinfo{year}{2020}).

\bibitem{Huang2022}
\bibinfo{author}{Huang, K.~S.}, \bibinfo{author}{Han, Z.},
  \bibinfo{author}{Kivelson, S.~A.} \& \bibinfo{author}{Yao, H.}
\newblock \bibinfo{title}{Pair-density-wave in the strong coupling limit of the
  holstein-hubbard model}.
\newblock \emph{\bibinfo{journal}{npj Quantum Materials}}
  \textbf{\bibinfo{volume}{7}}, \bibinfo{pages}{17} (\bibinfo{year}{2022}).

\bibitem{PhysRevLett.129.177001}
\bibinfo{author}{Wietek, A.}
\newblock \bibinfo{title}{Fragmented cooper pair condensation in striped
  superconductors}.
\newblock \emph{\bibinfo{journal}{Phys. Rev. Lett.}}
  \textbf{\bibinfo{volume}{129}}, \bibinfo{pages}{177001}
  (\bibinfo{year}{2022}).

\bibitem{PhysRevLett.74.3652}
\bibinfo{author}{Zhang, S.}, \bibinfo{author}{Carlson, J.} \&
  \bibinfo{author}{Gubernatis, J.~E.}
\newblock \bibinfo{title}{Constrained path quantum monte carlo method for
  fermion ground states}.
\newblock \emph{\bibinfo{journal}{Phys. Rev. Lett.}}
  \textbf{\bibinfo{volume}{74}}, \bibinfo{pages}{3652--3655}
  (\bibinfo{year}{1995}).

\bibitem{GuGu2020}
\bibinfo{author}{Gu, Y.}, \bibinfo{author}{Zhu, S.}, \bibinfo{author}{Wang,
  X.}, \bibinfo{author}{Hu, J.} \& \bibinfo{author}{Chen, H.}
\newblock \bibinfo{title}{A substantial hybridization between correlated
  $\mathrm{Ni}$-$d$ orbital and itinerant electrons in infinite-layer
  nickelates}.
\newblock \emph{\bibinfo{journal}{Communications Physics}}
  \textbf{\bibinfo{volume}{3}}, \bibinfo{pages}{84} (\bibinfo{year}{2020}).

\bibitem{science.aal5304}
\bibinfo{author}{Jiang, H.-C.} \& \bibinfo{author}{Devereaux, T.~P.}
\newblock \bibinfo{title}{Superconductivity in the doped hubbard model and its
  interplay with next-nearest hopping}.
\newblock \emph{\bibinfo{journal}{Science}} \textbf{\bibinfo{volume}{365}},
  \bibinfo{pages}{1424--1428} (\bibinfo{year}{2019}).

\bibitem{SANTOS2003}
\bibinfo{author}{Santos, R. R.~d.}
\newblock \bibinfo{title}{{Introduction to quantum Monte Carlo simulations for
  fermionic systems}}.
\newblock \emph{\bibinfo{journal}{{Braz. J. Phys.}}}
  \textbf{\bibinfo{volume}{33}}, \bibinfo{pages}{36 -- 54}
  (\bibinfo{year}{2003}).

\end{thebibliography}

\end{document}